# Multi-fidelity topology optimization of flow boiling heat transfer in microchannels

Yi Yuan[a], Li Chen[a, *], Qirui Yang[a], Lingran Gu[a], Wen-Quan Tao[a]

a: Key Laboratory of Thermo-Fluid Science and Engineering of MOE, School of Energy and Power Engineering, Xi'an Jiaotong University, Xi'an, Shaanxi 710049, China

* Corresponding author: Li Chen Email: lichennht08@mail.xjtu.edu.cn

**Abstract:** Topology optimization (TO) is a powerful method to design innovative structures with improved heat transfer performance. In the present study, a multi-fidelity TO method with a delicately defined objective function is developed for flow boiling heat transfer in microchannels. Low-fidelity TO is conducted for the reduced-order process of single-phase laminar convective heat transfer, which generates a set of structure candidates for subsequent high-fidelity evaluation of flow boiling heat transfer. To avoid the possible iteration between the low-fidelity TO and high-fidelity evaluation which leads to inefficient solution of the multi-fidelity TO, distributions of velocity, temperature and two-phase in microchannels with single-phase and/or flow boiling heat transfer are investigated and compared in detail, based on which a new objective function is delicately defined, which can be employed in the low-fidelity TO yet can stand for the performance of the high-fidelity problem. With the help of the new objective function, the efficiency of the multi-fidelity TO is significantly improved and TO structures are designed with hot spots eliminated, thermal resistance reduced and temperature uniformity improved. The present work provides a new method for TO of complicated heat and mass transfer problems.

## 1. Introduction

The increasing power of electronic devices brings challenge for heat dissipation. Among different heat dissipation methods, the microchannel heat sink in which forced convective heat transfer occurs is a promising technique with high heat transfer coefficient. In 1981, Tuckerman and Pease designed a parallel straight microchannel with liquid water as coolant [1], which can dissipate heat flux as high as 790 $W \cdot cm^{-2}$ with temperature rise of less than 71°C. Nowadays, the microchannel heat sink has been widely used in many fields such as cooling of chips, fuel cells and batteries. In the literature, great efforts have been devoted to the optimization of the microchannel heat sink in terms of geometric design, solid material and fluid type. The geometry design of the microchannels is very important as it not only determines the heat transfer area but also affects the fluid flow and heat transfer inside the microchannels. Different sizes [2-4], cross-section (rectangular, trapezoidal, circular, triangular shape, etc.) [5], channel patterns (straight, wavy, zigzag, serpentine, root or leaf shaped, spider-web shaped, etc.) [6] and fins/cavities [7] have been proposed and studied in the literature.

Compared with existing size, shape and/or bioinspired optimization of the microchannels, the topology optimization (TO) has the highest design freedom, which can generate optimized structures beyond the experience of designers. TO is a computational design method which automatically generates structures with optimum performance under formulated design conditions. The density-based method proposed by Bendsøe and Kikuchi [8] is one of the widely used methods for TO, which optimizes the distribution of materials in the design domain, and gives the relationship between design variables and material properties. Initially TO was adopted for designing mechanical structures, which was extended to other fields such as fluid flow problems [9, 10]. Borrvall and Petersson [11] first employed TO for fluid flow problem. The momentum equation was modified by adding a resistance term related to the design variable, and the resistance equals zero and infinity in the pure fluid and pure solid regions, respectively. Koga et al. [12] adopted TO for dual-objective optimization of convective heat transfer with constant heat sources based on the single-layer reduced-order model, where density filtering function was employed to avoid the checkerboard problem. There have been many subsequent studies on the optimization of microchannel convective heat transfer problems using TO. Zeng et al. [13] further proposed a two-layer model for convective heat transfer in microchannels, with one-layer for the flow channel and the other for the bottom solid substrate. Yan et al. [14] developed a new two-layer model based on the assumption of fourth-order polynomial temperature distribution in the fluid channel layer and linear temperature distribution in the solid substrate layer. Ozguc et al. [15] used a homogenization TO method to design the microchannel heat sink which consists of micro pin fin, and the size of micro pin fin was studied as an important parameter. Their experimental results showed that this interesting method is not only very friendly to additive manufacturing, but also does not deviate from the numerical simulation results. Their another study [16] proposed a flow-shifting TO heat sink with multiple inlets for cooling the chip with multiple different nonuniform thermal loads. Their experimental results showed that the flow-shifting TO heat sink reduces the thermal resistance at both operating loads by 10.7% and 6.8% relative to a baseline structure that does not design the inlets and flow paths for the specific heat loads. This TO design method for realistically varying thermal loads is very inspiring.

Currently, the TO studies of convective heat transfer in microchannels are mainly focusing on single-phase flow. Flow boiling can further enhance the heat transfer due to the latent heat related to phase change [17, 18]. Current research on flow boiling in conventional microchannels mostly focused on the design of microstructures for enhancing heat transfer. Li et al. [19] transformed the boundary layer structure by arranging micro pin fin fences on both sides of the channel to extend the evaporation region of the cooling fluid. Their results showed that this design achieves satisfactory results, with a 170% increase in heat transfer capacity over conventional microchannels. In fact, flow boiling heat transfer in microchannels has been extensively studied both numerically and experimentally in the literature [20-25]. However, currently there have been no studies directly applying the TO to flow boiling heat transfer processes. On the one hand, the flow boiling heat transfer processes present the features of evolving two-phase phases as well as other key variables such as velocity, pressure and temperature. Such an unsteady process with variations of multiple variables poses great challenging for the TO. On the other hand, even if TO of such processes achieves converged solution, which usually requires high computational resources and is easy to diverge, the solution has a high risk to be trapped in a locally optimal solution, as schematically shown in Fig. 1(a). In fact, Yaji et al. pointed out that for complex flow problems such as turbulent flow and/or phase-change heat transfer, strong multimodal distribution of the objective functions exists, and the TO of such problems is easy to be trapped in a locally optimal solution, or even a convergent solution cannot be achieved [26]. Therefore, in the literature, for complicated physical problems, researchers have to employ some reduced-order models to conduct TO. For example, Darcy equation was employed for the TO of turbulent flow and natural convective heat transfer [27, 28].

Considering the challenging of directly conducting TO for complicated physical problems, Yaji et al. [26] developed a multi-fidelity TO method for turbulent heat transfer in microchannels, in which there are two key steps, the low-fidelity TO for laminar convective heat transfer to generate a set of seeding structures, and the high-fidelity evaluation of the seeding structures by performing modeling of turbulent convective flow in these seeding structures to identify the best structure. TO microchannels with improved turbulent heat transfer performance were generated. Note that several seeding parameters are required to be predefined by implementing the low-fidelity TO. In their recent study [29], human selection of seeding parameters was replaced by an evolutionary algorithm which provided crossover and mutation structure generation to find high-performance structures in the high-fidelity solution space more efficiently.

In the present study, for the first time the above multi-fidelity TO scheme is employed for flow boiling heat transfer problems. The low-fidelity TO for single-phase convective heat transfer in microchannels is employed to generate seeding structures, with seeding parameters as different inlet pressures and different TO models. Then the seeding structures are evaluated by directly performing flow boiling heat transfer modeling in these seeding structures. The main contribution and innovation of the multi-fidelity TO study in the present study is that an new objective function which can stand for the performance of flow boiling heat transfer is proposed and employed in the low-fidelity TO, which is very important for improving the efficiency of multi-fidelity TO and avoiding the possible iteration between low-fidelity TO and high-fidelity evaluation. The remainder of the present paper is organized as follows. In Section 2, the total framework of the multi-fidelity TO is proposed. In Section 3, the low-fidelity TO model and the high-fidelity evaluation model are introduced. In Section 4, the multi-fidelity TO is conducted to improve the flow boiling heat transfer in microchannels. The new objective function is proposed which greatly improves the efficiency of the multi-fidelity TO. Finally, the conclusions are drawn in Section 5.

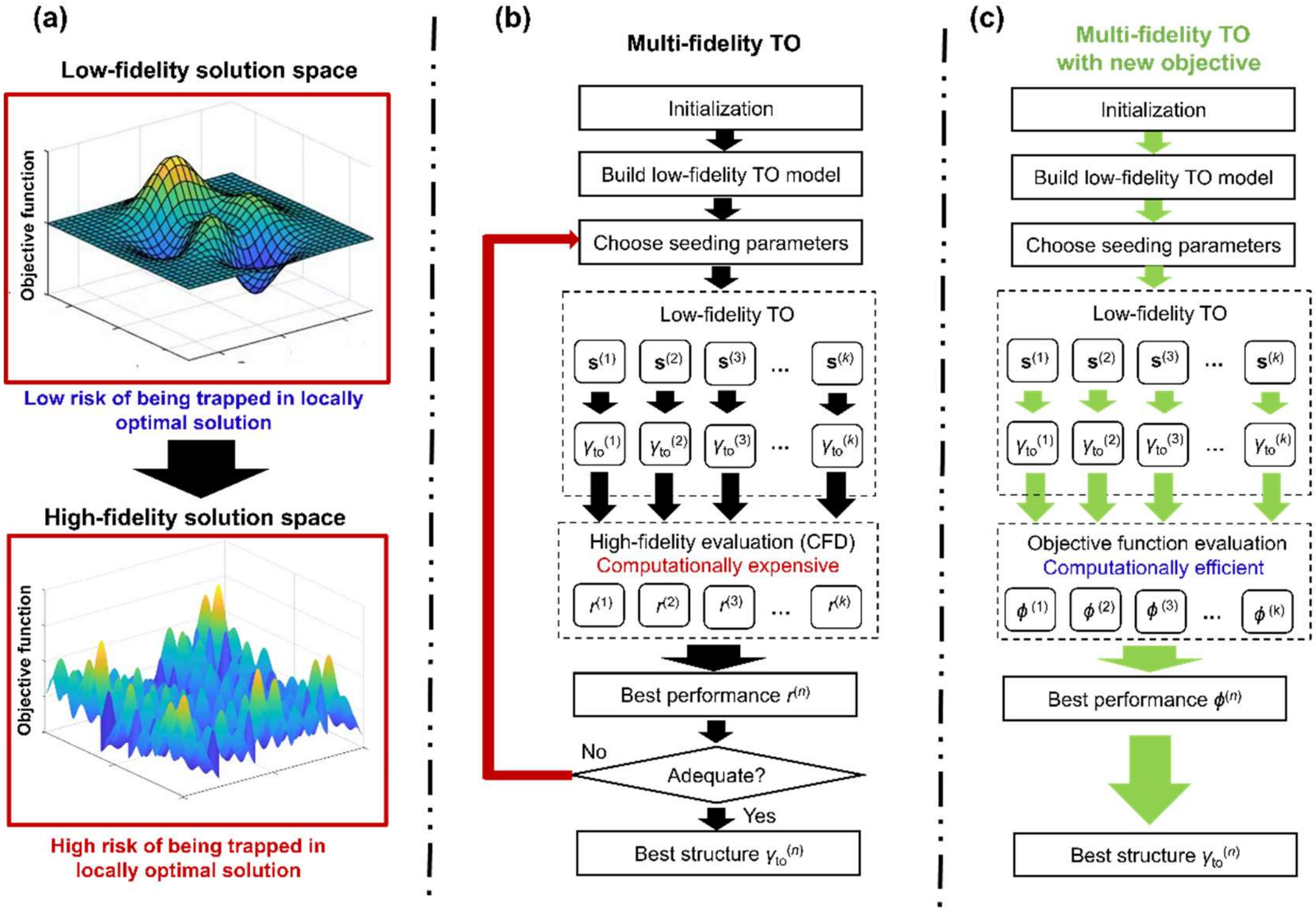

*Fig. 1. Framework of multi-fidelity TO. (a) Comparison of the solution space of the low-fidelity and high-fidelity TO. (b) Framework of the traditional multi-fidelity TO. (c) Multi-fidelity TO with new objective function.*

## 2. Framework of the multi-fidelity TO

The multi-fidelity TO method is helpful for the structural optimization of complicated physical problems [26], which consists of two key steps namely the low-fidelity TO and the high-fidelity evaluation as schematically shown in Fig. 1 (b). The risk of being trapped in a locally optimal solution for the low-fidelity TO is much lower than the direct TO of the original complex physical problems, as schematically shown in Fig. 1(a). In the low-fidelity TO, the original complicated problem under investigation is reduced to a relatively simple problem, for example, turbulent flow to laminar flow, or multi-phase flow to single-phase flow in the present study. The TO related to the reduced-order problem rather than the original complex problem is conducted to generate a set of structures. In the high-fidelity evaluation step, the structures generated by the low-fidelity TO, called seeding structures, are then studied, in which the original complex problem is solved and the performance is evaluated to choose the best structure from the seeding structures. From the above description, it can be found that the purpose of the low-fidelity TO is to provide structure candidates or seeding structures for the high-fidelity evaluation. Therefore, several seeding parameters will be defined for performing the low-fidelity TO, and the seeding parameters can be any parameters that would affect the TO generated structures. For example, by choosing different values of Reynolds number (*Re*), one can obtain different TO structures.

Based on the above description, the mathematical description of the multi-fidelity TO is introduced as follows. For the low-fidelity TO

$$\text{Low-fidelity TO} \quad \begin{cases} \text{minimize} & \tilde{J}(\gamma^{(j)}, \mathbf{s}^{(j)}) \\ \text{subject to} & \tilde{G}_i(\gamma^{(j)}, \mathbf{s}^{(j)}) \le 0, \; i=1,\cdots,M, \\ & 0 \le \gamma^{(j)} \le 1 \\ \text{for given} & \mathbf{s}^{(j)}, j=1,\cdots,K. \end{cases} \tag{1}$$

where $\tilde{J}$ and $\tilde{G}$ are the objective and constraint functions of the low-fidelity optimization problem, respectively. $\gamma$ is the design variable to describe the structure during the TO process. A node with $\gamma$=1 and $\gamma$=0 is fluid and solid, respectively. $i$ and $M$ are the index and number of constraint functions, respectively. $\mathbf{s}$ is the set of seeding parameters. $j$ expresses the sample point of $\mathbf{s}$ and $K$ is the number of sample points. Therefore, total $K$ cases of the low-fidelity optimization will be simulated, generating $K$ candidates or seeding structures by the TO for subsequent high-fidelity evaluation, with each candidate as $\gamma_{\text{to}}^{(j)}$.

The high-fidelity evaluation is formulated as follows

$$\text{High-fidelity evaluation} \quad \begin{cases} \text{minimize} \;\; J(\gamma_{\text{to}}^{(j)}) \\ \text{subject to} \;\; G(\gamma_{\text{to}}^{(j)}) \le 0, \; i{=}1,\cdots,M, \end{cases} \tag{2}$$

where $J$ and $G$ are the objective and constraint functions of the high-fidelity evaluation, respectively.

From the above description, it can be found that since the physical process solved in the low-fidelity TO is a reduced model of the original relatively complex physical problem, it is possible that the structures obtained by the low-fidelity TO present poor performance when evaluated in the framework of the high-fidelity problem. Therefore, one has to reformulate the low-fidelity problem to search for more seeding structures [26]. It is possible that several iterations are required between the low-fidelity TO and the high-fidelity evaluation as shown in Fig. 1(b), leading to inefficient solution of such multi-fidelity TO method.

In the present study, we try to in the framework of low-fidelity TO find an objective function $\tilde{J}$ in Eq. (1) that can represent the performance of the original complex physical problem. In this way, it is hoped that the seeding structures generated can pass the high-fidelity evaluation successfully and thus avoid the iterations between low-fidelity TO and high-fidelity evaluation mentioned above, as schematically shown in Fig. 1(c). The following question then naturally arises, how to define such an objective function? This is the main problem to be solved in the present study.

The method proposed in the present study is as follows. During the high-fidelity evaluation of the seeding structures, the distributions of key variables such as velocity, temperature and vapor phase will be analyzed in detail, and the main factors affecting the performance of the original physical problems will be identified. Then, the main factors will be projected to the low-fidelity TO by formulating a new objective function that can stand for or quantitatively describe these factors. As will be demonstrated by the results in Section 4, by employing such a new objective function, the low-fidelity

TO will generate structures with satisfactory performance in the high-fidelity evaluation.

Based on the above idea, the framework of the multi-fidelity TO in the present study is displayed in Fig. 1(c) and includes six steps. 1) initializing the parameter required for the high-fidelity modeling; 2) defining the low-fidelity TO problem; 3) choosing the seeding parameters; 4) conducting the low-fidelity TO to generate different seeding structures; 5) evaluating the seeding structures based on the corresponding objective function; 6) obtaining the seeding structure with the highest performance. From the above description, it can be found that Step 5 is very important to avoid the possible iteration between low-fidelity TO and high-fidelity evaluation and to improve the multi-fidelity TO efficiency.

## 3. Multi-fidelity TO of flow boiling heat transfer in microchannel heat sink

In the present study, flow boiling heat transfer in microchannels will be studied as shown in Fig. 2(a). The microchannel heat sink consists of three layers: the solid substrate layer, the flow channel layer, and the adiabatic solid top layer arranged in the *y*-direction. The cooling water enters the microchannels from the inlet, exchanges heat with the bottom surface and exits through the outlet. The goal of the present study is to improve the flow boiling heat transfer in the microchannel heat sink.

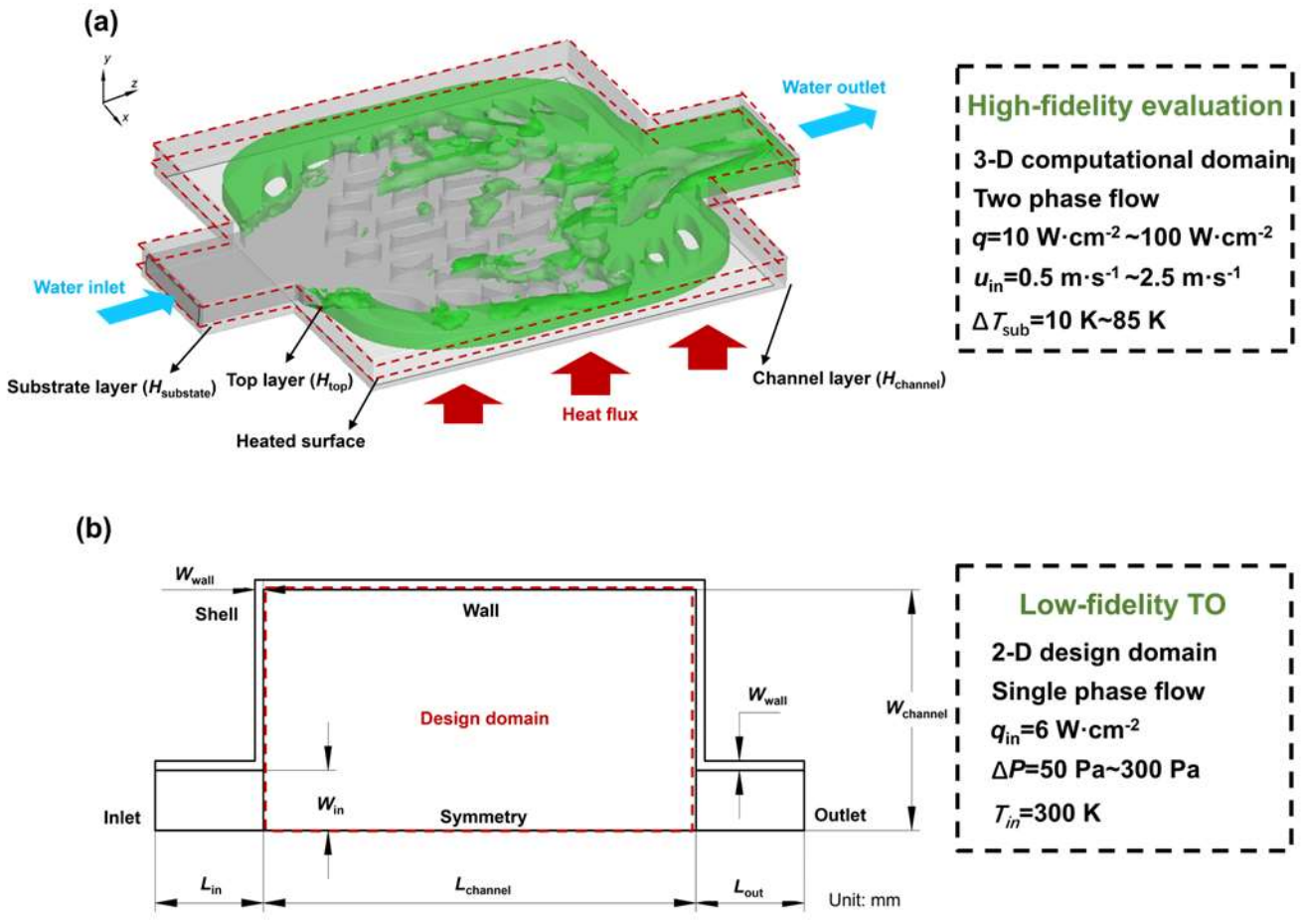

*Fig. 2. Schematic of the computational domain. (a) Schematic of the 3D heat sink. (a) Schematic of TO design domain.*

Directly conducting TO of such dynamic and coupled sub-processes of fluid flow, heat transfer and phase change is extremely difficult, if possible, due to dynamic evolution of the multiphase interface, many variables involved, and strong nonlinear characteristics. Therefore, the single-phase laminar heat transfer process in the microchannel is formulated as the low-fidelity problem as shown in Fig. 2(b), which is described in Section 3.1. Then in Section 3.2, the the phase change heat transfer mode for implementing the high-fidelity evaluation is introduced.

### 3.1 Low-fidelity topology optimization

To perform TO of the above laminar convective heat transfer in microchannels, in the literature several TO models have been proposed, including single-layer model [30], two-layer model [13] and even three-layer model [31]. The microchannel structures generated by the above TO models have been demonstrated to overwhelm traditional microchannels. Note that the TO structures generated by different TO models are not the same, and sometimes even present distinct characteristics. For example, the TO structures generated by the single-layer TO model usually have relatively large blocks compared with that generated by the two-layer models [12, 13], as shown in the literature as well as in Fig. 3 in the present study. Among the two-layer models, the one proposed by Yan et al. [14] employed a fourth-order polynomial temperature profile in the fluid layer and a linear temperature profile in the solid substrate layer, leading to very interesting TO structures which are different from previous two-layer model. It can be found that different TO models generate different TO structures, and this offers an opportunity to generate different seeding structures. In the present study, one single-layer TO model and two two-layer TO models will be employed, which are briefly introduced as follows and for more details one can refer to the Supplementary Material.

#### 3.1.1 2L model I ($s_1$=1)

For the two-layer TO model, both the fluid channel layer and the solid substrate model are simulated, and they are coupled by an additional source term in the energy conservation equation representing the heat exchange between the two layers. The first two-layer TO model employed, named 2L model I ($s_1$=1), is proposed by Yan et al. [14], for which the mass and momentum conservation equations in the fluid channel layer are as follows

$$\nabla \cdot u = 0 \tag{3}$$

$$\frac{6}{7}\rho_{\mathrm{fl}}(u \cdot \nabla)u = -\nabla p + \mu \nabla^2 u + F \tag{4}$$

where $u$, $\rho_{\mathrm{fl}}$, $p$, $\mu$ are the velocity, density, pressure and viscosity of the fluid, respectively. Note that Eq. (4) is different from the original Navier-Stokes equation because the 3D fluid flow in microchannels is reduced to 2D, and by simplifying the flow in the 3D channel to a parabolic shaped Poiseuille flow, a rescaling constant of 6/7 is added to the momentum equation [14]. $F$ is the Brinkman body force term [11]

$$F = -\alpha(\gamma)u \tag{5}$$

where $\alpha(\gamma)$ is the inverse permeability, which varies in fluid and solid regions due to the continuously changing material distributions during the TO process.

The energy conservation equations in both the fluid channel layer and the substrate layer are expressed as

$$\frac{2}{3}u \cdot \nabla\left(\rho_{\mathrm{fl}} c_{p,\mathrm{fl}} T_{\mathrm{c}}\right) = \frac{49}{52}\nabla \cdot \left(\lambda(\gamma)\nabla T_{\mathrm{c}}\right) + \frac{h(\gamma)(T_{\mathrm{b}} - T_{\mathrm{c}})}{2H_{\mathrm{c}}} \tag{6}$$

$$\nabla \cdot \left(\frac{\lambda_{\mathrm{s}}}{2}\nabla T_{\mathrm{b}}\right) + \frac{q_{\mathrm{in}}}{2H_{\mathrm{b}}} - \frac{h(\gamma)(T_{\mathrm{b}} - T_{\mathrm{c}})}{2H_{\mathrm{b}}} = 0 \tag{7}$$

where $c_{\mathrm{p,fl}}$ is the fluid heat capacity; $T_{\mathrm{c}}$ and $T_{\mathrm{b}}$ are the temperature of the channel layer and the solid substrate layer, respectively; $\lambda(\gamma)$ is the effective thermal conductivity; $\lambda_{\mathrm{s}}$ is the thermal conductivity of solid; $H_{\mathrm{c}}$ and $H_{\mathrm{b}}$ are the half height of the fluid channel and solid substrate layer; $q_{\mathrm{in}}$ is the heat flux applied at the bottom of the solid substrate layer. The heat transfer coefficient between the two layers $h(\gamma)$, the effective thermal conductivity $\lambda(\gamma)$, and the detailed deviation of Eqs. (6) and (7) are introduced in the Supplementary Material S.1.1.

### 3.1.2 2L model II ($s_1$=2)

The mass and the momentum conservation equations of the second kind of two-layer model, named 2L model II ($s_1$=2), are similar to Eqs. (3) and (4) [13]. The difference between the two 2L models is that the rescaling constant is not included in the conservation equations of 2L model II

$$\nabla \cdot u = 0 \tag{8}$$

$$\rho_{\mathrm{fl}}(u \cdot \nabla)u = -\nabla p + \mu \nabla^2 u + F \tag{9}$$

$$u \cdot \nabla\left(\rho_{\mathrm{fl}} c_{p,\mathrm{fl}} T_{\mathrm{c}}\right) = \nabla \cdot \left(\lambda(\gamma) \nabla T_{\mathrm{c}}\right) + \frac{h(\gamma)\left(T_{\mathrm{b}} - T_{\mathrm{c}}\right)}{2H_{\mathrm{c}}} \tag{10}$$

$$\nabla \cdot \left(\lambda_{\mathrm{s}} \nabla T_{\mathrm{s}}\right) + \frac{q_{\mathrm{in}}}{2H_{\mathrm{b}}} - \frac{h(\gamma)\left(T_{\mathrm{b}} - T_{\mathrm{c}}\right)}{2H_{\mathrm{b}}} = 0 \tag{11}$$

Details of the 2L model II can be found in the Supplementary Material S.1.2.

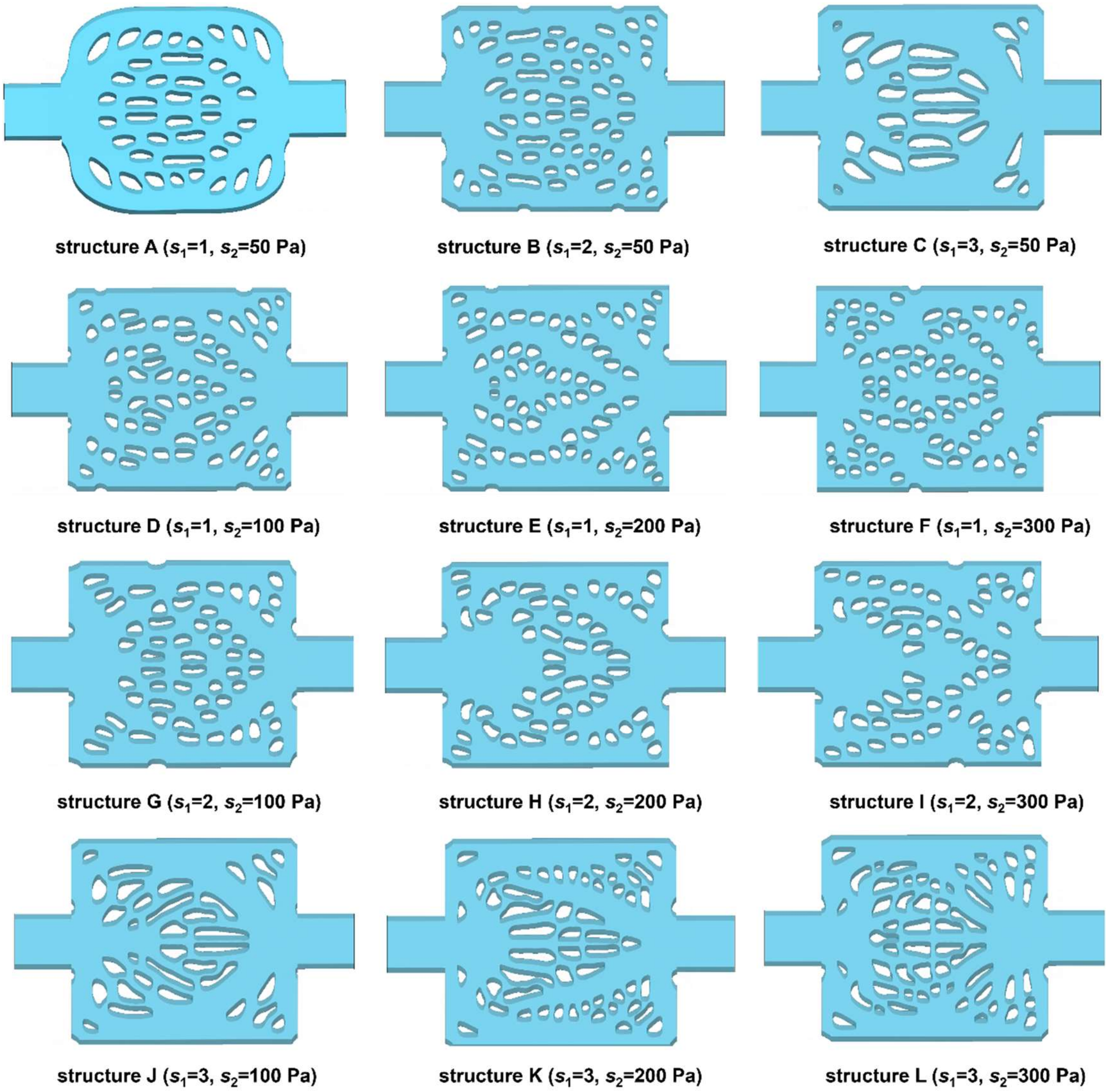

*Fig. 3. Structures generated by low-fidelity TO with different seeding parameters. $s_1$ is for different TO models and $s_2$ is for different inlet pressure.*

### 3.1.3 1L model ($s_1$=3)

Unlike the two-layer TO models above, in the one-layer model only the fluid channel is considered, and the heat flux at the bottom surface is directly added into the energy conservation equation as a source term. The governing equations in the one-layer model, called 1L model ($s_1$=3) are as follows [30]

$$\nabla \cdot u = 0 \tag{12}$$

$$\rho_{\mathrm{fl}}(u\cdot\nabla)u = -\nabla p + \mu\nabla^2 u + F \tag{13}$$

$$\gamma\rho_{\mathrm{fl}}c_{p\mathrm{fl}}\left(u\cdot\nabla\right)T_{\mathrm{c}} = [(1-\gamma)\frac{\lambda_{\mathrm{s}}}{\lambda_{\mathrm{f}}} + \gamma]\nabla^2 T_c + (1-\gamma)\frac{q_{\mathrm{in}}}{2H_{\mathrm{c}}} \tag{14}$$

The 1L model is introduced in detail in the Supplementary Material S.1.3.

### 3.1.4 Model implementation

Based on the governing equations with different $s_1$, the low-fidelity TO problem for the microchannel heat sink is defined as

$$\begin{aligned} &\min: \; T_{\max} \\ &\text{s.t.}: \begin{cases} \text{governing equations} \\ \dfrac{\int_{\Omega_{\mathrm{d}}} \gamma dA}{A_{\Omega_{\mathrm{d}}}} = f_{\mathrm{c}} \\ 0 \leqslant \gamma \leqslant 1 \end{cases} \end{aligned} \tag{15}$$

where $T_{\max}$ is the maximum temperature of the heat sink, and the minimum $T_{\max}$ corresponds to the highest heat transfer performance. $A_{\Omega\mathrm{d}}$ is the area of the design domain $\Omega_{\mathrm{d}}$. $f_{\mathrm{c}}$ is the fluid volume fraction constraint.

To obtain a differentiable maximum operator, the maximum temperature in the computational domain is calculated by a generalized $p$-norm

$$T_{\max} = \left(\frac{1}{N}\sum\nolimits_N T_i^p\right)^{\frac{1}{p}} \tag{16}$$

where $N$ is the total number of nodes in the computational domain and $p$ is the norm parameter set as 10 [32, 33].

In the present study, the inverse permeability $\alpha(\gamma)$ in Eq. (5) is interpolated by the Darcy interpolation function [34]

$$\alpha\left(\gamma\right) = \alpha_{\mathrm{fl}} + \left(\alpha_{\mathrm{s}} - \alpha_{\mathrm{fl}}\right)\frac{1-\gamma}{1+q_{\mathrm{f}}\gamma} \tag{17}$$

where $\alpha_{\mathrm{fl}}$ is the inverse permeability in the fluid domain and $\alpha_{\mathrm{s}}$ is the inverse permeability in the solid domain [34], and the values of $\alpha_{\mathrm{s}}$ and $\alpha_{\mathrm{fl}}$ for the above three models can be found in the Supplementary Material S.1.4. $q_f$ is the penalization parameter to regulate the convexity of the interpolation function. In the present study, $q_f$ is set as 1000, 500, 100 and 20 in different TO steps. In each step, the TO solver runs 50 times or meets the tolerance of 0.01.

The effective thermal conductivity $\lambda(\gamma)$ is interpolated by the rational approximation of material properties (RAMP) function [35]. To avoid the checkerboard problems, the Helmholtz based on partial differential equation (PDE) density filter is adopted [36, 37]. The density filter may lead to the intermediate density area at the fluid-solid interface. To obtain clearer fluid-solid interface, the hyperbolic tangent projection is employed [38]. The interpolation, density filter and the projection models with the values of key parameters can be found in the Supplementary Material S.1.4 and Table. 1.

#### 3.1.5 Boundary and initial conditions

During the TO process in the present study, the design domain is half of the 2D heat sink, as shown in Fig. 2(b). The working fluid is liquid water. The pressure boundary condition is applied at the inlet with different $p_{in}$. Different values of $p_{in}$ also serve as a seeding parameter for low-fidelity TO ($s_2$). Constant fluid inlet temperature is applied at the inlet with $T_{in}$ of 300 K. Constant heat flux boundary condition is applied to the bottom surface with $q_{in}$ and for other walls the adiabatic condition is applied. For all the inner fluid-solid interface, the conjugate boundary condition is employed. Non-slip condition is applied to all walls. The symmetric boundary condition is applied to the symmetric axis of the heat sink. Initially, $\gamma$ is set as 0.8. All the detailed geometric and operating parameters during the TO process are listed in Table. 1.

#### 3.1.6 Numerical procedures

In the present study, the low-fidelity TO is solved in the COMSOL Multiphysics 6.0. The process is divided into five steps. The computational domain is discretized and initialized, and the seeding parameters ($s_1$ corresponding to the three TO models and $s_2$ for different inlet pressure) are defined. The governing equations are solved based on the finite element method (FEM). The sensitivity analysis is performed by the continuous adjoint method. The design variable $\gamma$ is updated by the moving asymptote method (GCMMA) [39]. Finally, the structure is adapted by filter and projection. The above steps are iterated repeatedly until the convergence criterion is satisfied.

The 2D TO structure obtained needs to be converted to the 3D structure. This is done as follows, first adding a 0.4 mm external solid boundary to the 2D structure. Then, the 2D structure is stretched in the $z$ direction with the height of the fluid channel layer. Finally, the stretched structure is assembled with substrate layer and top layer.

Based on the above low-fidelity TO and post-processing process, different 3D optimized structures are obtained with three values of $s_1$ and three values of $s_2$, as shown in Fig. 3. It can be found that the two two-layer models ($s_1$=1 and $s_1$=2) generate solid blocks relatively smaller compared with that generated by the one-layer model ($s_1$=3). The 12 structures in Fig. 3 generated by the low-fidelity TO will be evaluated in terms of performance of flow boiling heat transfer, and the model is introduced in Section 3.2.

## 3.2 High-fidelity evaluation

### 3.2.1 Flow boiling heat transfer model

The flow boiling heat transfer process in the heat sink is numerically studied by the phase-change two-phase flow model [40]. Only the key equations, including conservation equations of mass, momentum, energy and volume fraction, are listed here, and the details can be found in the Supplementary Material S.2.

$$\frac{\partial(\rho)}{\partial t} + \nabla\cdot(\rho\mathbf{u}) = 0 \tag{18}$$

$$\frac{\partial(\rho\mathbf{u})}{\partial t} + \nabla\cdot(\rho\mathbf{u}\mathbf{u}) = -\nabla p + \nabla\cdot[\mu(\nabla\mathbf{u} + \nabla\mathbf{u}^{\mathrm{T}})] + \rho\mathbf{g} + \mathbf{F} \tag{19}$$

$$\frac{\partial(\rho c_{\mathrm{p}} T)}{\partial t} + \nabla\cdot(\rho c_{\mathrm{p}} T\mathbf{u}) = \nabla\cdot(\lambda\nabla T) + S_{\mathrm{E}} \tag{20}$$

$$\frac{\partial(\rho_k C_k)}{\partial t} + \nabla\cdot(\rho_k C_k\mathbf{u}) = \dot{m}_k \tag{21}$$

where $t$, $p$, $\mathbf{u}$, $\mathbf{g}$, $T$, $\rho$, $\mu$, $c_{\mathrm{p}}$, $\lambda$, is the time, pressure, velocity, gravitational acceleration, temperature, density, dynamic viscosity, heat capacity and thermal conductivity in the framework of two-phase flow, respectively. The fluid properties such as $\rho$, $\mu$ and $\lambda$ are the volume-averaged values, while $c_{\mathrm{p}}$ is mass averaged value. The subscript $k$ corresponds to the $k^{th}$ phase. Volume of Fluid (VOF) method is adopted for two-phase flow, and $C$ is the volume fraction of different phases. $C_1$ and $C_2$ are defined as the volume fraction of the primary phase (water) and secondary phase (vapor). The body force term $\mathbf{F}$ is calculated by the continuum surface force (CSF) model [41] which converts surface tension force at the phase interface to the body force. $S_E$ in Eq. (20) is the energy source term caused by the phase change, which can be calculated based on the phase-change rate $\dot{m}$

$$S_E = h_{fg}\dot{m} \tag{22}$$

where $h_{fg}$ is the latent heat. $\dot{m}$ can be used to calculate the liquid reduction rate $\dot{m}_1$, vapor generation rate $\dot{m}_2$

$$\dot{m} = \dot{m}_1 = -\dot{m}_2 \tag{23}$$

Details of the calculation of $\dot{m}$ can be found in the Supplementary Material S.2.3

### 3.2.2 Initial and boundary conditions

In the high-fidelity evaluation, the liquid water flows into the heat sink from the inlet with constant velocity of $u_{in}$ and temperature as $T_{in}$ with subcooling ($\Delta T_{sub} = T_{sat} - T_{in}$). The outflow boundary condition is adopted for the outlet. A constant heat flux $q$ is applied at the bottom surface of the substrate layer. No-slip boundary and fluid-solid coupling heat transfer conditions are used for all the fluid-solid interfaces. Adiabatic boundary condition is used for the remaining external solid wall. The reference pressure is set as 101325 Pa. As the initial conditions, the absolute pressure, temperature and velocity in the whole computational domain are set as 101325 Pa, $T_{in}$ and zero, respectively. Furthermore, the physical properties of the fluid and solid during the flow boiling simulation are listed in Table. 1.

### 3.2.3 Numerical procedures

In the present study, the flow boiling heat transfer is solved by FLUENT 2022 R2. The pressure-implicit with the splitting of operators (PISO) scheme is adopted to solve the unsteady laminar incompressible flow. The Geo-reconstruct interface reconstruction as well as with user-defined functions (UDF) is coupled with the VOF method to build and identify the two-phase interface during the flow boiling process. The phase-change model introduced in Section 3.2.1 is integrated with the VOF method by UDF. The second-order upwind scheme is used to discrete the convective terms. During the solving process, the adaptive time step is used to ensure that the Courant number (Co) is less than 0.8.

In our previous work, the phase-change model has been validated by a film boiling process [40], and for the purpose of brevity, the validation result is described in the Supplementary Material S.3. Moreover, to obtain the mesh independence solution, Structure A in Fig. 3 is discretized by different numbers of meshes (197726, 452755, 635590 and 817494). The results show that the average temperature of the bottom surface ($T_{ave}$) with 635590 cells is only 0.08 K larger than that with 817494 cells, which is thus selected to balance the solution accuracy and the computational resources.

## 4. Results and discussion

In this section, the structures generated by the TO of the single-phase heat transfer (low-fidelity problem) in Section 3.1 as shown in Fig. 3 will be evaluated by the flow boiling heat transfer model (high-fidelity evaluation) introduced in Section 3.2. The vapor and temperature distributions will be discussed in detail. Particularly, the present study tends to propose a new objective function which both can be calculated under low-fidelity conditions and can represent the performance of flow boiling heat transfer. Furthermore, efforts are devoted to directly employ this new objective function in the low-fidelity TO to generate TO structures with improved flow boiling heat transfer performance.

### 4.1 Flow boiling heat transfer in the TO structures

In this section, flow boiling heat transfer processes in the seeding structures generated by low-fidelity TO are explored under different heat flux (10~100 $W \cdot cm^{-2}$), inlet coolant velocity (0.5~2.5 $m \cdot s^{-1}$) and subcooling (10~85 K). Note that the above values are commonly studied in the literature [42-44]. Without loss of generality, the first structure in Fig. 3 with $s_1$=1 and $s_2$=50 Pa is chosen as an example. Using the flow boiling heat transfer model in Section 3.2, the dynamic evolution of vapor phase, the velocity, pressure and temperature fields can be predicted, based on which the relationship between heat flux $q$ and thermal resistance $r$ can be obtained

$$r = \frac{T_{\mathrm{ave}} - T_{\mathrm{sat}}}{q} \tag{24}$$

with $T_{\mathrm{ave}}$ as the area-weighted average temperature at the solid bottom surface when the boiling is stabilized. Note that there are still periodic temperature fluctuations when the boiling is considered to be stabilized, therefore $T_{\mathrm{ave}}$ is taken as the highest average value during the boiling.

First, effects of $q$ on $r$ are studied as plotted in Fig. 4(a) with inlet coolant velocity as 0.5 m $s^{-1}$ and the subcooling temperature as 10 K. As shown in Fig. 4(a), for $q$ lower than 50 W·$cm^{-2}$, $r$ gradually decreases as $q$ increases due to the occurrence of phase change. After that, $r$ increases indicating poorer heat transfer performance as the bottom surface of the channel is gradually covered by the vapor generated [40]. The heat flux at 50 W·$cm^{-2}$ is called the turning point in the flow boiling curve. Before this point, as the flow boiling becomes intense, the heat transfer becomes stronger. After this point, a large amount of vapor accumulates in the microchannels which is difficult to discharge and covers the solid surface, causing the heat exchange to deteriorate sharply.

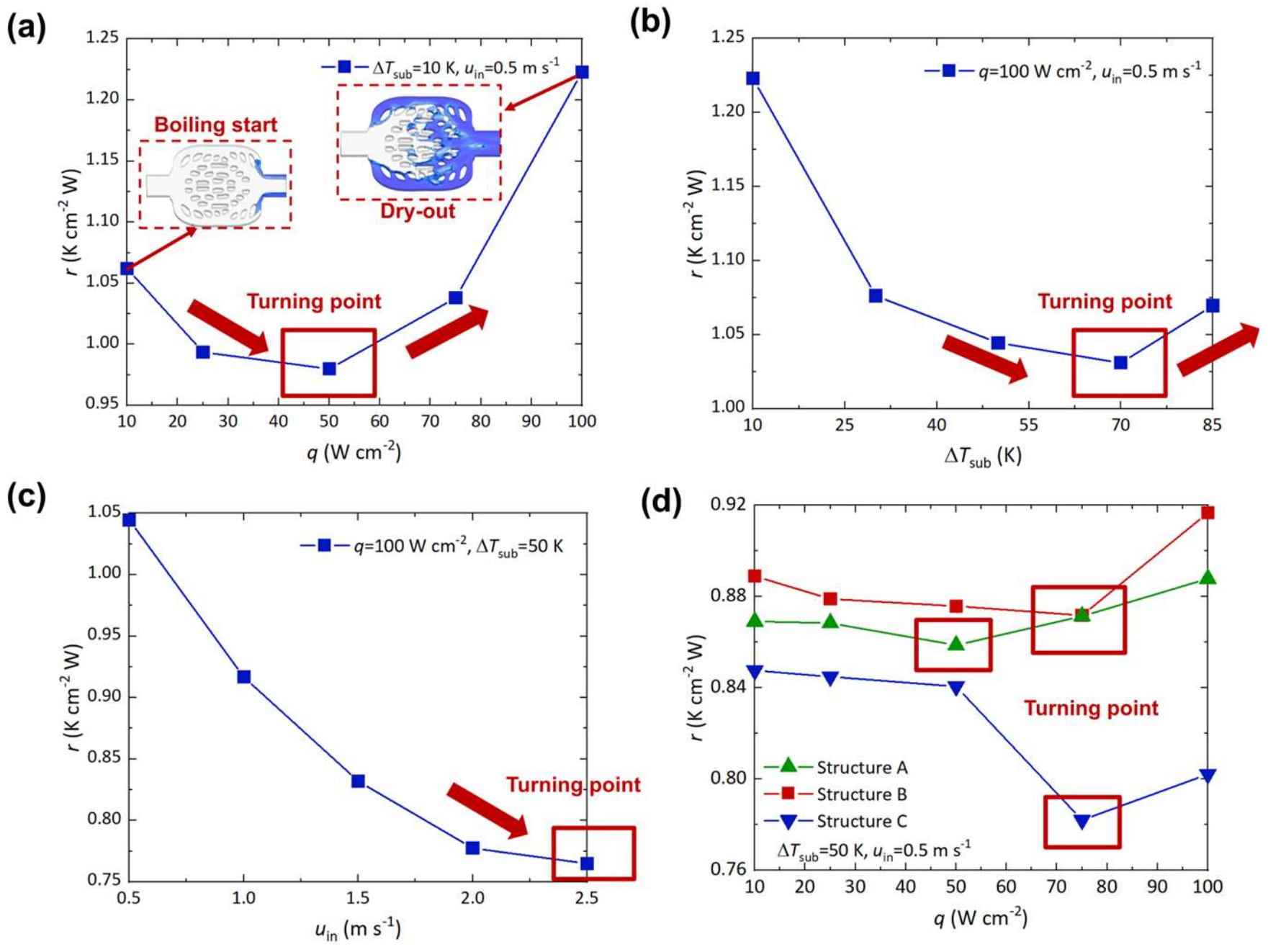

*Fig. 4 Flow boiling heat transfer performance under different parameters. (a) r-q curve (structure A in Fig. 3). (b) r-$\Delta T_{sub}$ curve (structure A). (c) r- $u_{in}$ curve (structure A). (d) r-q curves for different structures.*

To more clearly understand the above results, the two-phase distributions and temperature fields at different times under different $q$ are provided in Fig. 5(a). Both the temperature field at $y = (H_b+H_c/2)$ and that at the solid bottom surface are provided. Under a low $q$ of 10 W·cm$^{-2}$, the cooling water is not heated to $T_{sat}$ until the outlet, and vapor is generated near the outlet. The temperature field shows that the overall temperature is low, and the high temperature regions are mainly in the four corners, while the center is with lower temperature. This is due to the fact that the velocity at the corners is relatively low, leading to poor local heat transfer and thus higher temperature.

The results under $q$ of 25 W·cm$^{-2}$ and 50 W·cm$^{-2}$ in Fig. 5(a) show that the overall temperature is higher and the flow boiling is more drastic due to higher heat flux. In the upstream and downstream corners, the temperature rise is the fastest, and the local temperature reaches $T_{sat}$ more quickly. As a result, a large amount of vapor is generated at the upstream and downstream corners, which moves along the side walls. However, the removal of vapor is slow because of low velocity at the corners. This causes the accumulation of vapor, which deteriorates the local heat transfer. As $q$ rises further to 75 W·cm$^{-2}$ and 100 W·cm$^{-2}$, the temperature of the center region also reaches $T_{sat}$, and then a large amount of vapor is generated, which converges and blocks the outlet. This dry-out phenomenon in the heat sink corresponds to the rise of heat resistance after the turning point in Fig. 4(a). In particular, the entire outlet region shows high temperature at the bottom of the heat sink in the dry-out state, which is different from cases with low $q$ of 10 W·cm$^{-2}$~50 W·cm$^{-2}$.

Effects of the subcooling with $\Delta T_{sub}$ of inlet coolant varied from 10 K to 85 K are also explored. Such range of $\Delta T_{sub}$ is commonly studied in experiments in the literature [45]. Besides, $q$ and $u_{in}$ are set constantly as 100 W·cm$^{-2}$ and 0.5 m·s$^{-1}$. The $r$-$\Delta T_{sub}$ relationship is shown in Fig. 4(b), and the two-phase distributions and temperature fields under different $\Delta T_{sub}$ are shown in Fig. 5(b). It can be found that $r$ decreases and then increases with increasing $\Delta T_{sub}$. Such trend can be clearly explained by the two-phase distribution in Fig. 5(b). It can be found that a large amount of vapor is generated at the center and the corner region with a low $\Delta T_{sub}$, which accumulates and blocks the outlet, so that the dry-out phenomenon occurs in microchannels. However, as $\Delta T_{sub}$ increases, the vapor generated is less as shown in Fig. 5(b). Therefore, less vapor blockage alleviates the dry-out phenomenon in microchannels, leading to lower $r$. As $\Delta T_{sub}$ increases further to 50 K and 70 K, $r$ decreases at a relatively slow rate because of the weakening of the boiling at the center. When $\Delta T_{sub.}$ is higher than 70 K, $r$ starts to increase, which means that vapor accumulation in the corners further deteriorates the heat transfer.

Finally, effects of different inlet velocity $u_{in}$ from 0.5 m·s$^{-1}$ to 2.5 m·s$^{-1}$ on the flow boiling heat transfer are investigated with $q$ and $\Delta T_{sub}$ fixed as 100 W·cm$^{-2}$ and 50 K. The $r$-$u_{in}$ relationship is shown in Fig. 4(c), and the two-phase distributions and temperature fields under different $\Delta T_{sub}$ are shown in Fig. 5(c). The results show that $r$ monotonously decreases as $u_{in}$ increases but the decreasing rate becomes smaller. The two-phase distribution in Fig. 5(c) shows that as velocity increases, the vapor at the outlet as well as at the downstream corner can be gradually removed, and correspondingly the local heat transfer can be enhanced. However, the alleviation of vapor accumulation at the upstream corner is insignificant even when the velocity is 2 m s$^{-1}$. This can be well explained by the streamline plotted in Fig. 5(c).

Based on the above parametric study, it can be found that the temperature at the upstream and downstream corners of the heat sink is higher due to the low local velocity, which further leads to the local accumulation of vapor and results in higher resistance for heat transfer from the solid bottom layer to the fluid. Therefore, if the local vapor accumulation can be alleviated or even eliminated, it is expected that the local heat transfer can be enhanced and the maximum temperature at the bottom surface can be reduced.

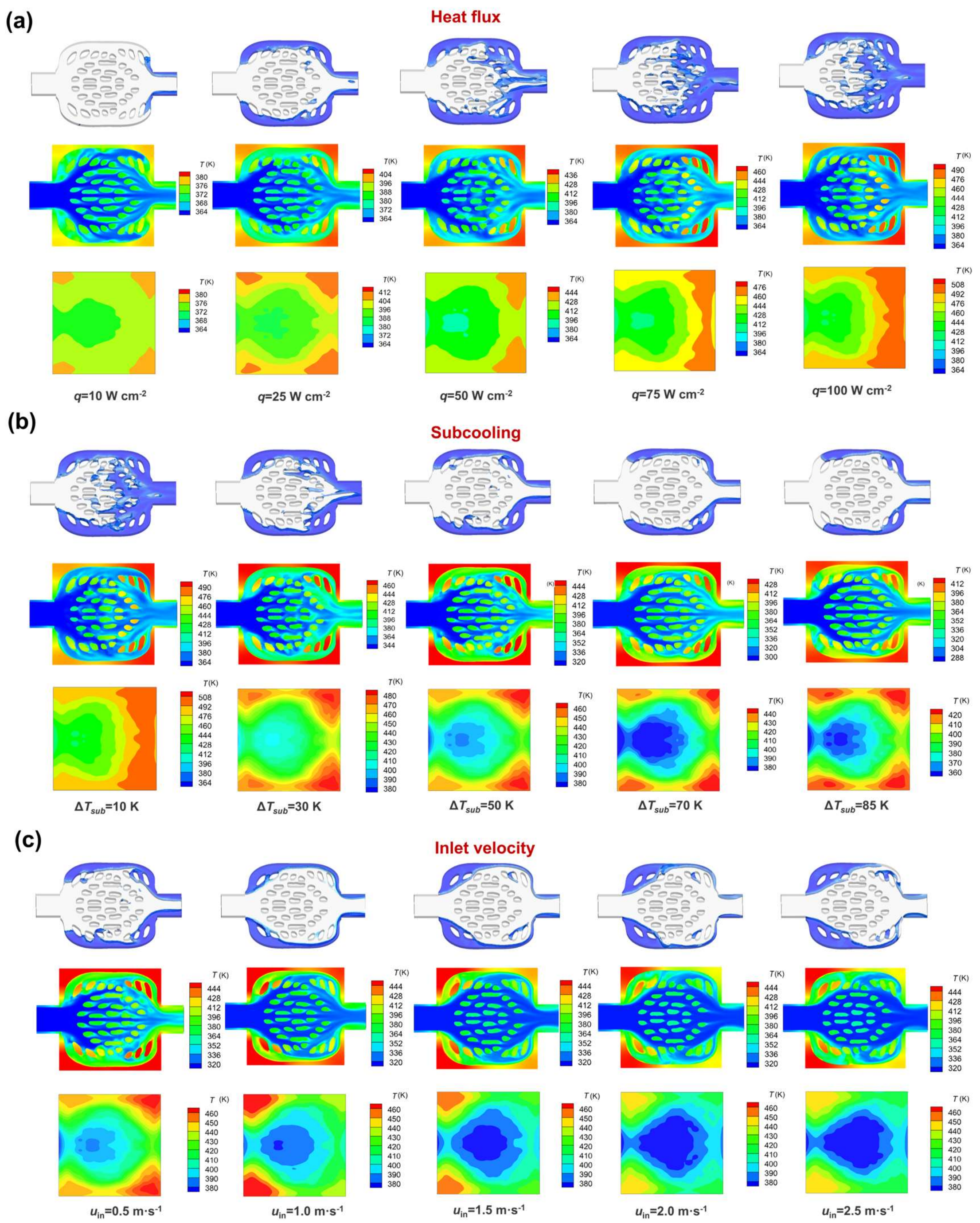

*Fig. 5. Results of high-fidelity evaluation under different operating parameters with structure A (Two-phase distributions, temperature fields at $y$ = ($H_b$+$H_c$/2) and temperature fields of bottom surface). Effects of (a) Heat flux, (b) Subcooling and (c) Inlet velocity.*

### 4.2 Comparison between low-fidelity and high-fidelity results

From the discussion in Section 4.1, it is inferred that if a TO structure generated by the low-fidelity structure can result in relatively strong convection at the corners of the microchannel heat sink, the local vapor accumulation can be eliminated. Therefore, here the temperature distributions of all the TO structures under low-fidelity conditions are displayed in Fig. 6, where the temperature is normalized by the highest temperature of all the structures. Roughly, it can be found that the structures can be divided into two categories according to the temperature distribution. For Structures A, B and G, the temperature at the corners is much higher than in other regions. The structures, including D, E, F, H, I and J, although the corner temperature is not that much higher than other regions compared with Structures A, B and G, also can be classified into the first category. For Structures C, K and L, quite different temperature fields can be observed, namely the corners are with lower temperature and in the entire domain the temperature is more uniform. Such distribution of C, K and L indicates that their corners can be well cooled by the coolant and the local heat transfer is relatively strong, which is demonstrated by the velocity fields of different structures in the Supplementary Material Fig. A4. Therefore, when coming to the high-fidelity process of flow boiling heat transfer, it is expected that the well-cooled corners with lower temperature will hinder the vapor accumulation and enhance the local phase-change heat transfer.

The maximum normalized temperature $T^*_{max}$ for each structure is also provided in Fig. 6. It can be found that $T^*_{max}$ of structures C, K and L is also relatively lower compared with other structures. Finally, the distribution uniformity of temperature is also evaluated by the following formula

$$S = \sqrt{\frac{\int (T^* - T^*_{ave})^2 \mathrm{d}A}{A}} \tag{25}$$

where $T^*_{ave}$ is the averaged normalized temperature. A smaller $S$ indicates more uniform temperature distribution. From Fig. 6, it also can be found that $S$ of Structures C, K and L are relatively lower than other structures. The lower $T^*_{max}$ and more uniform temperature distribution in the low-fidelity problem also indicate that the risk for local hot spot in the high-fidelity problem would be lower.

To prove the above inference, flow boiling heat transfer in all the TO structures in Fig. 3 is studied and evaluated. Fig. 4(d) and Fig. 7 selectively display the results for Structures A, B and C under different $q$ with $\Delta T_{sub}$ and $u_{in}$ as 50 K and 1 m·s$^{-1}$. For each curve in Fig. 4(d), it can be found that the variation of $r$ with $q$ is similar to that in Fig. 4(a). The results in Figs. 7(a)~(c) also show that as $q$ increases more vapor is generated and then covers the corners in each structure. It is worth noting that $r$-$q$ curve of Structure C undergoes a sharp drop from 50 to 75 W·cm$^{-2}$, which can be well explained by the results in Fig. 7(c) as for Structure C there is no phase change before 75 W·cm$^{-2}$. In fact by comparing the $r$-$q$ curves of Structures A-C, the results show that the thermal resistance of Structure C is the lowest, and that of Structure B is the highest. The results in Fig. 7(b) show that the vapor accumulation already exists in Structure B under $q$ of 50 W·cm$^{-2}$. Furthermore, it can be found that the thermal resistance of Structure A is slightly lower than Structure B. This can be well explained by the images in Figs. 7(a) and (c). For example, at $q$ of 50 W·cm$^{-2}$, vapor only accumulates in the upstream corner in Structure A, while in Structure B vapor film can be observed along the sides walls from the upstream to the downstream.

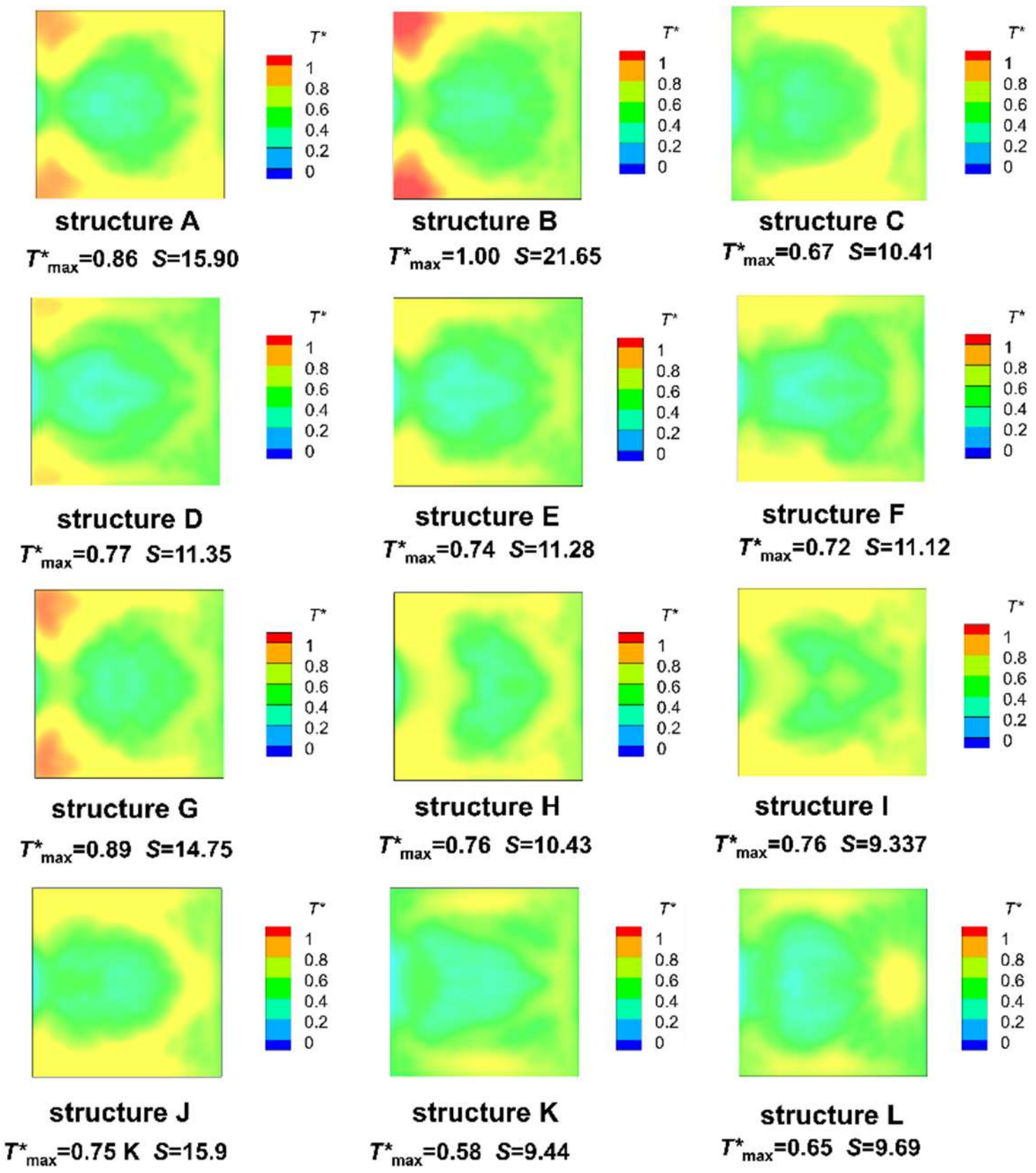

*Fig. 6. Normalized temperature fields, maximum normalized temperature and temperature uniformity of different structures under the low-fidelity condition (Single-phase laminar convective heat transfer).*

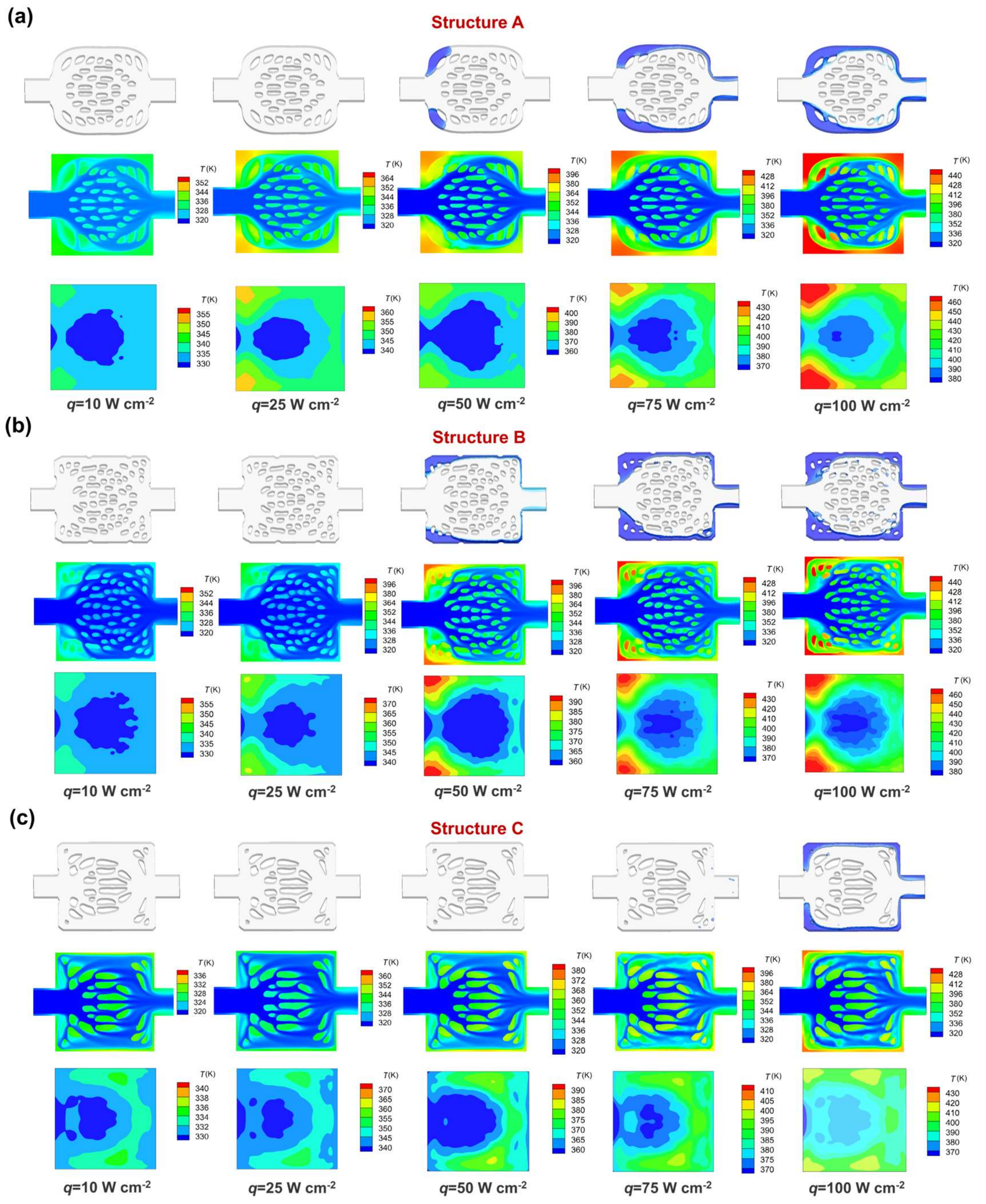

*Fig. 7. Results of high-fidelity evaluation under different q with structure A, B and C (Two-phase distributions, temperature fields at $y=(H_b+H_c/2)$ and temperature fields of bottom surface). (a) Structure A. (b) Structure B. (c) Structure C.*

Further, Fig. 8 shows the distribution of vapor and temperature for all the TO structures with $q$, $\Delta T_{sub}$ and $u_{in}$ as 100 W·cm$^{-2}$, 50 K and 1 m·s$^{-1}$. It also can be found that the temperature characteristics of Structures K and L in Fig. 6 are similar to that in Fig. 8, namely hot spot is eliminated in the corners, maximum temperature is lower and temperature distribution is more uniform. Therefore, comparing the results in Figs. 6~8, it can be concluded that a TO structure obtained by the low-fidelity simulation with local hot spots eliminated and more uniform temperature distribution can result in good performance of flow boiling heat transfer. Such conclusion is crucial for the proposal of new objective function for low-fidelity optimization, and will be discussed in the next section.

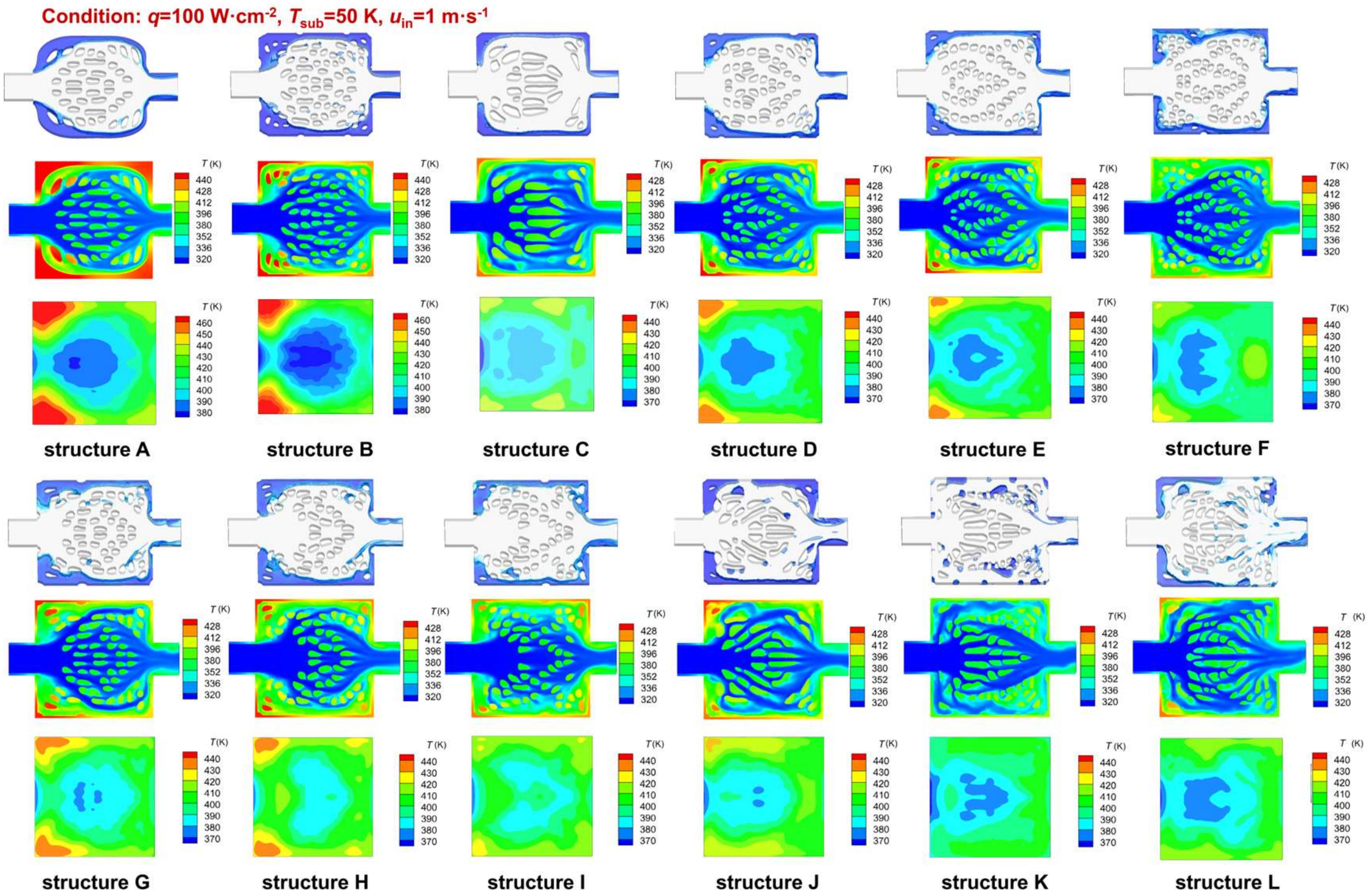

*Fig. 8. Results of high-fidelity evaluation with all structures (Two-phase distributions, temperature fields at $y=(H_b+H_c/2)$ and temperature fields of bottom surface).*

### 4.3 New objective function for multi-fidelity topology optimization

Note that in Eq. (15), the objective for the low-fidelity TO is to minimize the maximum temperature, or the lowest thermal resistance. Fig. 9(a) displays the thermal resistance for the 12 structures in Fig. 3 related to single-phase convective heat transfer. Corresponding to the results in Fig. 6, it can be found that Structure A is with the highest thermal resistance while Structure L is with the lowest. For comparisons, Fig. 9(b) plots the thermal resistance related to the flow boiling heat transfer with the condition provided in the image. Interestingly, the result in Figs. 9(a) and (b) is not one-to-one match. For example, Structure A is the worst in Fig. 9(a) but the worst structure in Fig. 9(b) is B. Structure C, while its performance is poor in Fig. 9(a), overwhelms most of structures in Fig. 9(b). The best structure is L in Fig. 9(a) but is K in Fig. 9(b).

The above results indicate that the objective function defined by Eq. (15) in the low-fidelity TO, namely minimizing the maximum temperature, cannot well describe the performance of flow boiling heat transfer. Therefore, a new objective function should be proposed, which can be employed in the low-fidelity TO and yet can represent the performance of high-fidelity problem. Based on the results and discussions in Sections 4.1 and 4.2, this new objective should take into account the following three aspects namely minimum maximum temperature, more uniform temperature distribution, and lowest corner temperature. Thus, the following formula is proposed

$$\phi = \frac{\int \frac{T - T_{\mathrm{in}}}{T_{\max} - T_{\mathrm{in}}} \frac{T}{T_{\max,\,\mathrm{ref}}} \frac{\Delta x}{\Delta x_{\mathrm{ref}}} dA}{A_{\mathrm{bot}}} \tag{26}$$

where the numerator is an integral over the entire bottom cells. $T_{\max}$ is the maximum temperature at the bottom surface of the current structure. $\Delta x$ is the distance from the current cell to the center of the computational domain. $\Delta x_{\mathrm{ref}}$ is the distance from the corner to the center of the computational domain as a reference value. $A_{\mathrm{bot}}$ is the area of the bottom surface. $T_{\max,\,\mathrm{ref}}$ is the maximum temperature of all the structures under low-fidelity conditions, which is a reference value to ensure that the objective function is dimensionless. There are three terms in the integral of the numerator. The first and second terms together characterize the overall temperature and temperature uniformity. The third term represents the weight factor of effects of every mesh in the domain, with the weight for the corner point as the highest. Eq. (26) indicates the corner temperature has the highest weight factor and its temperature should be as low as possible to avoid local vapor accumulation. One may argue that the vapor may not always accumulate at the corners, and for a flow boiling heat transfer in other structures rather than the microchannel studied in the present study, the distribution of temperature and vapor may be quite different. Under such circumstance, the location with high risk of high temperature (or vapor accumulation) will also be assigned a high weight factor by defining proper weight functions.

To demonstrate if the new objective can describe the performance of the TO structures for flow boiling heat transfer, the relationship between this new objective function and the thermal resistance of different structures for the flow boiling heat transfer is shown in Fig. 9(c). The results in Fig. 9(c) are exciting. It can be found that the best Structure K has the lowest $\phi$ of all the structures and the worst Structure B has the highest $\phi$. Besides, for all other structures, the objective function and the thermal resistance are correlated, clearly demonstrating that the new objective function can well describe the performance of the flow boiling heat transfer in the TO structures. As illustrated in Fig. 9(d), the objective function can be employed under low-fidelity conditions, and then predict the performance under high-fidelity conditions. In our study, other forms of objective function rather than Eq. (26) such as $(T\text{-}\ T_{\mathrm{in}})^2/(T_{\mathrm{max}}\text{-}T_{\mathrm{in}})^2(\Delta x/\Delta x_{\mathrm{ref}})$ were also tested, but the established mapping between the objective function and high-fidelity thermal resistance is not as good as Eq. (26).

The next step is to employ this new objective function in the low-fidelity TO to find out if it can generate TO structures with good performance in the high-fidelity evaluation as well explored in Section 4.4.

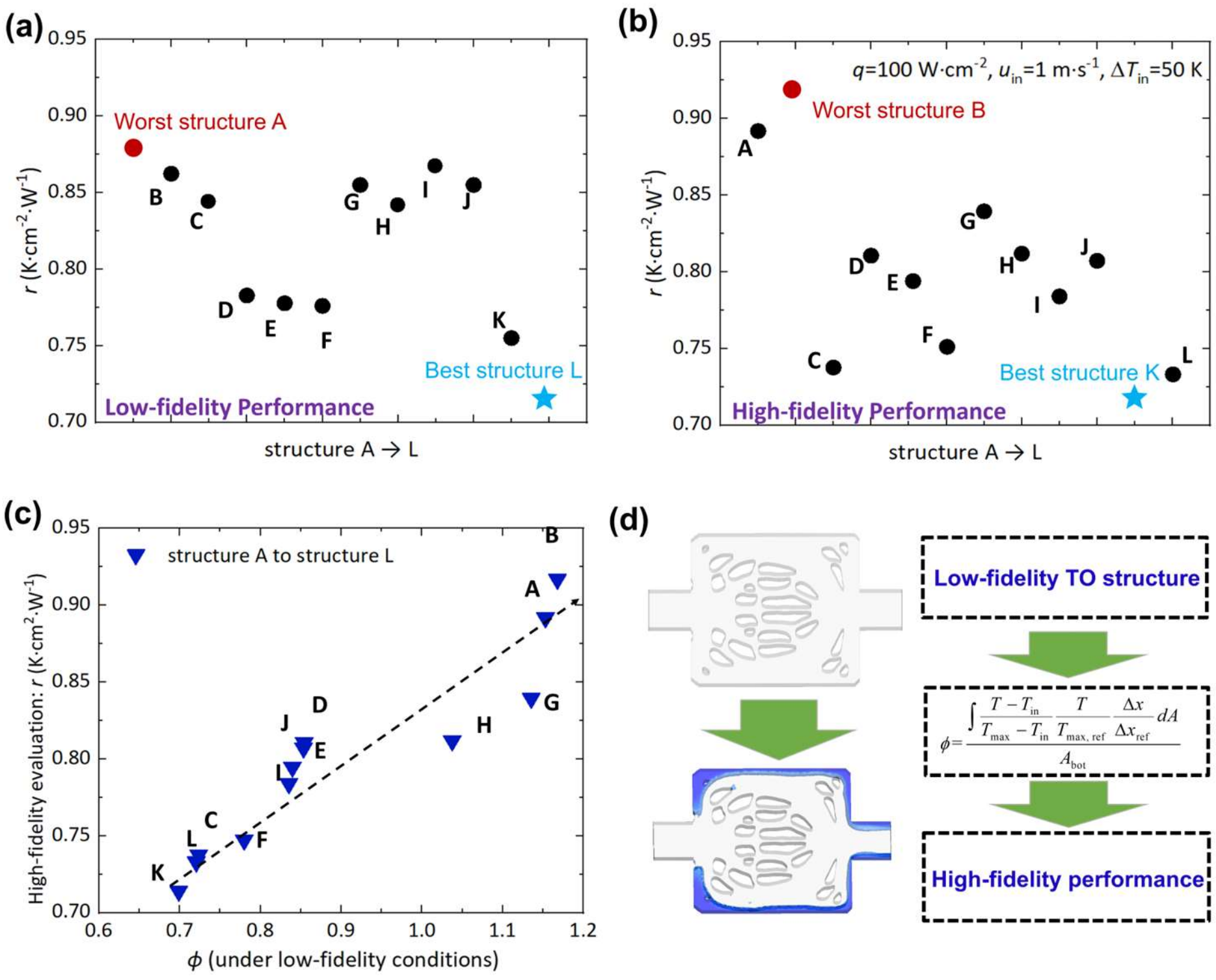

*Fig. 9. Validation of the new objective function. (a) r for different structures under low-fidelity conditions. (b) r for different structures under high-fidelity conditions. (c) Relationship between r and $\phi$. (d) Schematic of the relationship between low-fidelity TO, new objective function, and high-fidelity performance.*

### 4.4 Multi-fidelity topology optimization based on the new objective function

In this section, the new objective function is adopted in the low-fidelity TO to generate new structures and the TO problem is redefined as follows

$$\begin{aligned}
&\min: \quad \phi=\frac{\int \frac{T-T_{\mathrm{in}}}{T_{\max}-T_{\mathrm{in}}} \frac{T}{T_{\max,\,\mathrm{ref}}} \frac{\Delta x}{\Delta x_{\mathrm{ref}}} dA}{A_{\mathrm{bot}}} \\
&\text{s.t.:} \quad \begin{cases} \text{governing equations} \\ \dfrac{\int_{\Omega_{\mathrm{d}}} \gamma dA}{A_{\Omega_{\mathrm{d}}}} = f_{\mathrm{c}} \\ 0 \leqslant \gamma \leqslant 1 \end{cases}
\end{aligned} \tag{27}$$

The TO structures with the above new objective functions as shown in Fig. 10(a) are generated with $s_1$=1~3 and $s_2$=50 and 300 Pa . It can be found that these new structures prefer to generate solids in the center of the design domain, which can direct the inlet water more efficiently to the corners and domain boundaries, thus enhancing the local convective heat transfer. The normalized temperature fields of these new structures are further shown in Fig. 10(b). The temperature fields show that $\phi_1$, $\phi_2$, $\phi_5$ and $\phi_6$ all avoid high temperature at the corners, which is similar to the high-performance Structures K and L. To further quantitatively characterize the temperature field of the new structures, $T^*_{\max}$ and $S$ of each new structure are also provided in Fig. 10(b). It can be found that the new structures all have low $T^*_{\max}$, which is lower than that of all the structures in Fig. 6 except Structure K. Furthermore, $S$ of Structures $\phi_1$, $\phi_2$, $\phi_5$ and $\phi_6$ is lower than that of all the structures in Fig. 6 except Structures I, K and L. The low maximum temperature and high temperature uniformity of the new structures clearly demonstrate the new objective function plays its role during the low-fidelity TO.

Now the above structures generated by the new objective function in the low-fidelity TO will be evaluated in the high-fidelity simulation, and the conditions are the same as that in Fig. 8. The results are displayed in Fig. 10(c). It can be found that in the new structures the vapor accumulation in the corners is more or less alleviated. In particular, there is no vapor accumulation in Structure $\phi_6$, which is because Structure $\phi_6$ has the tightest central solid arrangement that directs most of the cooling fluid to the edge and corner regions.

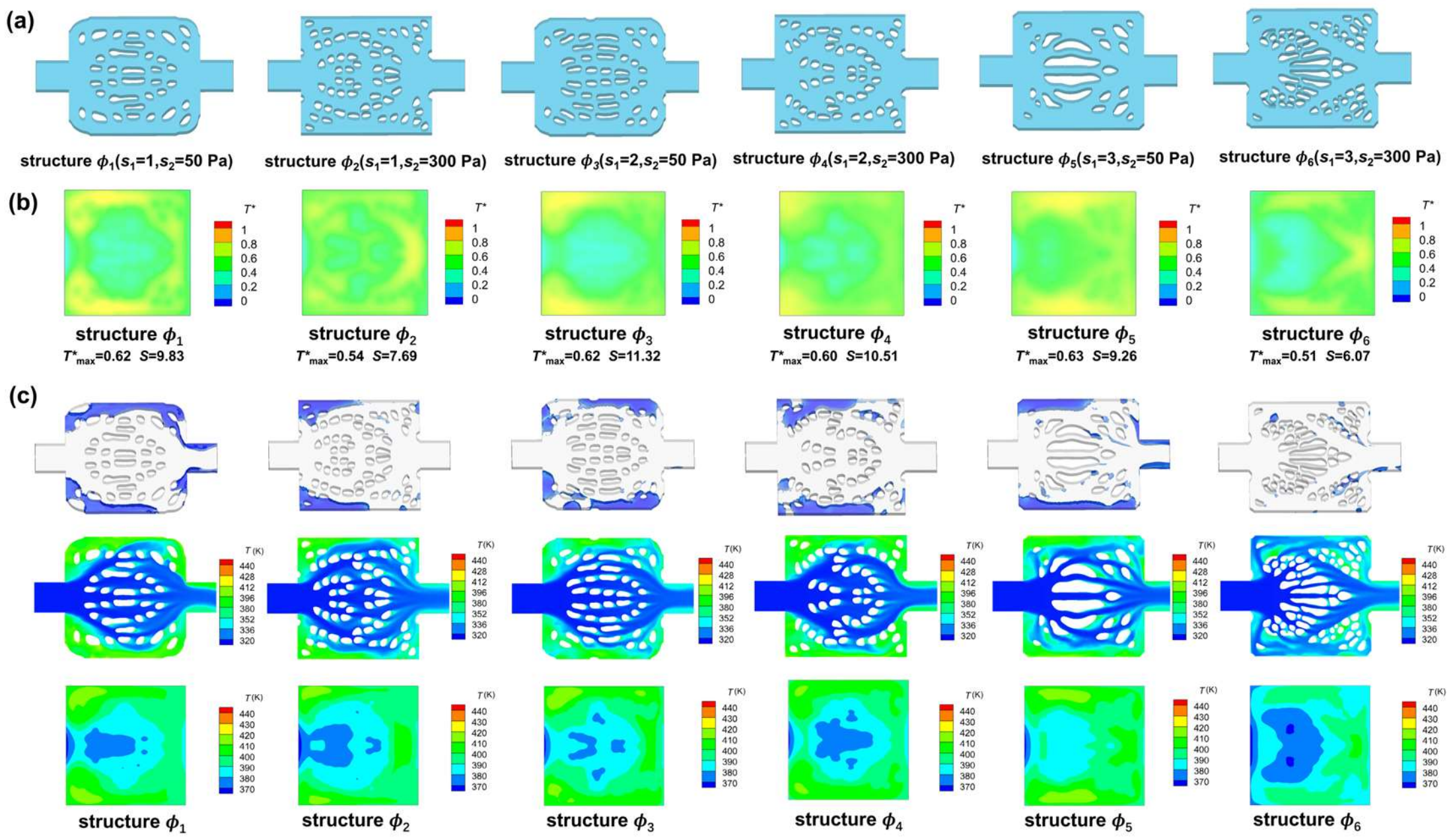

*Fig. 10. The TO results based on the new objective function. (a) 2D TO structure. (b) Normalized temperature fields, maximum normalized temperature and temperature uniformity of different structures under the low-fidelity condition. (c) Two-phase distribution, temperature fields at $y=(H_b+H_c/2)$ and temperature field of the bottom surface.*

Finally, the new generated TO structures are comprehensively evaluated in terms of the thermal resistance, the highest temperature and the temperature distribution uniformity, which are also compared with the best and worst Structure B and K in Fig. 9. The results of thermal resistance in Fig. 11(a) show that the six new generated structures all have relatively low thermal resistance, much lower than Structure B. Among them, Structure $\phi_6$ leads to the lowest thermal resistance which is even 12% lower than that of Structure K. Moreover, the high-fidelity performance regions of different objective functions are marked in Fig. 11(a), and it can be noticed that the difference between maximum and minimum $r$ of the TO structure generated based on the new objective function Eq. (26) is much lower than that of the TO structure generated based on the original objective function Eq. (15), although $r$ of some new structures is still higher than the Structure K. This indicates that the new objective function makes it easier to obtain structures with high flow boiling heat transfer performance without the need for multiple generation and selection as shown in Fig.

1(b). Without a deliberate choice of seeding parameters, the new structures already outperform most of the structures generated by the original objective function.

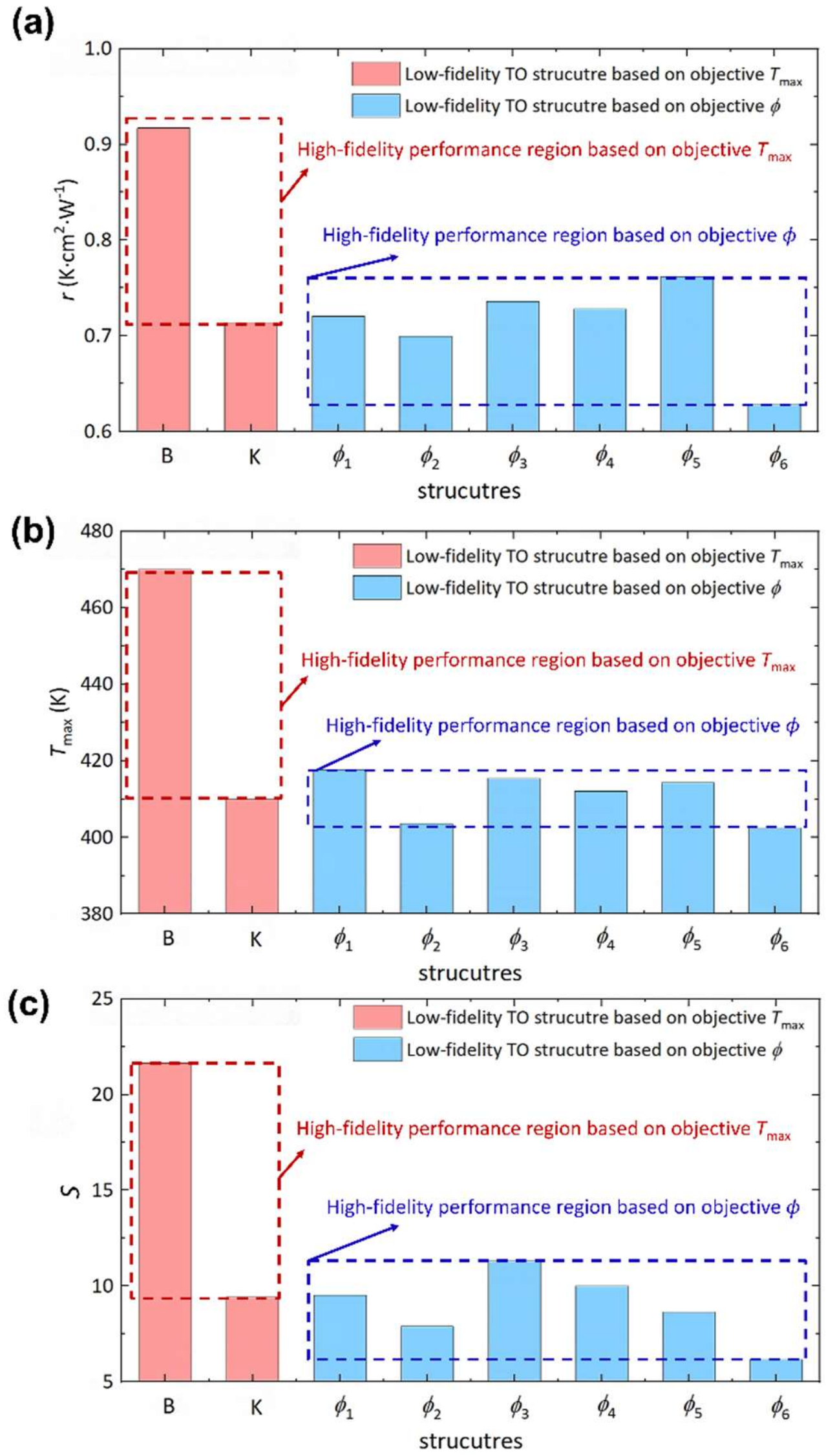

*Fig. 11. Comprehensive comparison between different TO structures. (a) Thermal resistance. (b) $T_{max}$. (c) Temperature distribution uniformity.*

The maximum temperature and temperature uniformity of the six new structures, Structure B and K under high-fidelity conditions are further calculated and shown in Figs. 11(b) and (c). The results show that the maximum temperature of the six new structures is significantly lower than that of Structure B, and the maximum temperature of Structures $\phi_2$ and $\phi_6$ is lower than that of Structure K. The results for $S$ are similar. Finally, the high-fidelity performance regions of different objective functions are marked in Figs. 11(b) and (c), and it can be found that for the maximum temperature and $S$, the difference between the different structures based on the new objective function is also small, again demonstrating that the new objective function alleviates the dependence on the seeding parameters compared to the original objective function.

## 5. Conclusions

Topology optimization is a powerful method to design microchannel heat sink with high heat transfer performance. However, existing TO methods are difficult to optimize complex flow heat transfer problems, such as flow boiling heat transfer and turbulent heat transfer. In this study, a multi-fidelity TO method with delicately defined objective function is developed for flow boiling heat transfer in microchannels. First, with different seeding parameters, low-fidelity TO structures are generated under the single-phase laminar convective heat transfer condition, which are adopted for the subsequent high-fidelity evaluation of flow boiling heat transfer. Second, in the high-fidelity evaluation, the two-phase distributions, temperature fields and velocity fields of these TO structures are studied under different flow boiling conditions. Third, by comparing the performance and physical fields of the same structure under low-fidelity and high-fidelity conditions, a new objective function is proposed, which can be employed in the low-fidelity TO yet can stand for the performance of the high-fidelity problem. Finally, in order to verify the performance of the new objective function, the low-fidelity TO based on the new objective function is performed, and the generated new structures are evaluated by flow boiling heat transfer modeling in terms of thermal resistance, maximum temperature and temperature uniformity, which show good performance. The main conclusions are as follows.

(1) Identifying the key factors that limit the flow boiling heat transfer in the seeding structures generated by the low-fidelity TO is crucial. In the present study, the low-velocity region at the microchannel corners which causes the local vapor accumulation is the major factor deteriorating the flow boiling heat transfer performance, and thus the low-velocity corner region should be eliminated.

(2) Comparing the distributions of key variables in the TO structures for the low-fidelity problem and the high-fidelity problem is important for proposing the new objective function. The new objective function proposed in the present study can reduce the maximum temperature, increase the temperature distribution uniformity and eliminate the local high temperature at the microchannel corners. The new objective function is well correlated to the performance of flow boiling heat transfer.

(3) Adopting the new objective function in the low-fidelity TO, which avoids the possible iterations between the low-fidelity TO and high-fidelity evaluation, can efficiently generate structures with enhanced flow boiling heat transfer performance. The new objective function also alleviates the dependence on seeding parameters and reduces the performance difference between different TO structures generated.

**Acknowledgement**

The authors acknowledge the support of National Natural Science Foundation of China (52376074).

***Table 1*** *Key parameters for low-fidelity TO and high-fidelity evaluation*

| Parameters | Symbol | Values |
|---|---|---|
| Key parameters for low-fidelity TO | | |
| Inlet half width | $W_{in}$ | 1.25 mm |
| Inlet length | $L_{in}$ | 2.5 mm |
| Channel half width | $W_{channel}$ | 5 mm |
| Channel length | $L_{channel}$ | 10 mm |
| Substrate layer height | $H_{substrate}$ | 0.2 mm |
| Channel layer height | $H_{channel}$ | 0.5 mm |
| Inlet temperature | $T_{in}$ | 300 K |
| Pressure drop | $\Delta p$ | 50~300 Pa |
| Bottom heat flux | $q_{in}$ | 6 W·cm$^{-2}$ |
| Key parameters for high-fidelity evaluation | | |
| Top layer height | $H_{top}$ | 0.2 mm |
| Wall width | $W_{\mathrm{wall}}$ | 0.2 mm |
| Density (liquid/vapor/solid) | $\rho$ | 998/0.554/2700 kg·m$^{-3}$ |
| Viscosity (liquid/vapor) | $\mu$ | 2.82×10$^{-4}$/1.26×10$^{-5}$ Pa·s |
| Specific heat capacity (liquid/vapor/solid) | $c_{\mathrm{p}}$ | 4.22/2.08/0.88 kJ·kg$^{-1}$·K$^{-1}$ |
| Thermal conductivity (liquid/vapor/solid) | $\lambda$ | 0.68/0.026/202.4 W·m$^{-1}$·K$^{-1}$ |
| Surface tension coefficient | $\sigma$ | 0.0648 N·m$^{-1}$ |
| Bottom heat flux | $q$ | 10~100 W·cm$^{-2}$ |
| Inlet subcooling | $\Delta T_{\mathrm{sub}}$ | 10~85 K |
| Inlet velocity | $u_{\mathrm{in}}$ | 0.5~2.5 m·s$^{-1}$ |

## Reference

1. Tuckerman, D.B. and R. Pease, *High-performance heat sinking for VLSI.* IEEE Electron Device Letters, 1981. **2**(5): p. 126-129.
2. Naphon, P., et al., *ANN, numerical and experimental analysis on the jet impingement nanofluids flow and heat transfer characteristics in the micro-channel heat sink.* International Journal of Heat and Mass Transfer, 2019. **131**(MAR.): p. 329-340.
3. Husain, A. and K.Y. Kim, *Shape Optimization of Micro-Channel Heat Sink for Micro-Electronic Cooling.* IEEE Transactions on Components and Packaging Technologies, 2008. **31**(2): p. 322-330.
4. Hadad, Y., et al., *Three-objective shape optimization and parametric study of a micro-channel heat sink with discrete non-uniform heat flux boundary conditions.* Applied Thermal Engineering, 2019.
5. Sadique, H. and Q. Murtaza, *Heat transfer augmentation in microchannel heat sink using secondary flows: A review.* International Journal of Heat and Mass Transfer, 2022. **194**: p. 123063.
6. Zheng, R., et al., *Development of a hierarchical microchannel heat sink with flow field reconstruction and low thermal resistance for high heat flux dissipation.* International Journal of Heat and Mass Transfer, 2022. **182**: p. 121925.
7. Rajalingam, A. and S. Chakraborty, *Effect of micro-structures in a microchannel heat sink–a comprehensive study.* International Journal of Heat and Mass Transfer, 2020. **154**: p. 119617.

8. Bendsøe, M.P. and N. Kikuchi, *Generating optimal topologies in structural design using a homogenization method.* Computer methods in applied mechanics and engineering, 1988. **71**(2): p. 197-224.
9. Liu, X., et al., *Topology optimization of the manifold microchannels with triple-objective functions.* Numerical Heat Transfer, Part B: Fundamentals, 2021. **80**(5-6): p. 89-114.
10. Luo, J.-W., et al., *Three-dimensional topology optimization of natural convection using double multiple-relaxation-time lattice Boltzmann method.* Applied Thermal Engineering, 2024. **236**: p. 121732.
11. Borrvall, T. and J. Petersson, *Topology optimization of fluids in Stokes flow.* International journal for numerical methods in fluids, 2003. **41**(1): p. 77-107.
12. Koga, A.A., et al., *Development of heat sink device by using topology optimization.* International Journal of Heat and Mass Transfer, 2013. **64**: p. 759-772.
13. Zeng, S. and P.S. Lee, *Topology optimization of liquid-cooled microchannel heat sinks: An experimental and numerical study.* International Journal of Heat and Mass Transfer, 2019. **142**: p. 118401.
14. Yan, S., et al., *Topology optimization of microchannel heat sinks using a two-layer model.* International Journal of Heat and Mass Transfer, 2019. **143**(NOV.): p. 118462.1-118462.16.
15. Ozguc, S., et al., *Experimental study of topology optimized, additively manufactured microchannel heat sinks designed using a homogenization approach.* International Journal of Heat and Mass Transfer, 2023. **209**: p. 124108.
16. Ozguc, S., L. Pan, and J.A. Weibel, *Topological optimization of flow-shifting microchannel heat sinks.* International Journal of Heat and Mass Transfer, 2023. **207**: p. 123933.
17. Garimella, S.V. and C.B. Sobhan, *Transport in microchannels-a critical review.* Annu.rev.heat Transf, 2003. **13**(13).
18. Li, W. and Z. Wu, *A general correlation for evaporative heat transfer in micro/mini-channels.* International Journal of Heat and Mass Transfer, 2010(9/10): p. 53.
19. Li, W., et al., *Supercapillary architecture-activated two-phase boundary layer structures for highly stable and efficient flow boiling heat transfer.* Advanced materials, 2020. **32**(2): p. 1905117.
20. Deng, D., L. Zeng, and W. Sun, *A review on flow boiling enhancement and fabrication of enhanced microchannels of microchannel heat sinks.* International Journal of Heat and Mass Transfer, 2021. **175**: p. 121332.
21. Prajapati, Y.K. and P. Bhandari, *Flow boiling instabilities in microchannels and their promising solutions–A review.* Experimental Thermal and Fluid Science, 2017. **88**: p. 576-593.
22. Kandlikar, S.G., *Fundamental issues related to flow boiling in minichannels and microchannels.* Experimental thermal and fluid science, 2002. **26**(2-4): p. 389-407.

23. Harirchian, T. and S.V. Garimella, *A comprehensive flow regime map for microchannel flow boiling with quantitative transition criteria.* International Journal of Heat and Mass Transfer, 2010. **53**(13-14): p. 2694-2702.
24. Xu, L. and J. Xu, *Nanofluid stabilizes and enhances convective boiling heat transfer in a single microchannel.* International Journal of Heat and Mass Transfer, 2012. **55**(21-22): p. 5673-5686.
25. Wang, G., P. Cheng, and H. Wu, *Unstable and stable flow boiling in parallel microchannels and in a single microchannel.* International Journal of Heat and Mass Transfer, 2007. **50**(21-22): p. 4297-4310.
26. Yaji, K., S. Yamasaki, and K. Fujita, *Multifidelity design guided by topology optimization.* Structural and Multidisciplinary Optimization, 2020. **61**: p. 1071-1085.
27. Zhao, X., et al., *A "poor man's approach" to topology optimization of cooling channels based on a Darcy flow model.* International Journal of Heat and Mass Transfer, 2018. **116**: p. 1108-1123.
28. Asmussen, J., et al., *A "poor man's" approach to topology optimization of natural convection problems.* Structural and Multidisciplinary Optimization, 2019. **59**(4): p. 1105-1124.
29. Yaji, K., S. Yamasaki, and K. Fujita, *Data-driven multifidelity topology design using a deep generative model: Application to forced convection heat transfer problems.* Computer Methods in Applied Mechanics and Engineering, 2022. **388**: p. 114284.
30. Matsumori, T., et al., *Topology optimization for fluid–thermal interaction problems under constant input power.* Structural and Multidisciplinary Optimization, 2013. **47**: p. 571-581.
31. Zhao, J., et al., *Topology optimization of planar cooling channels using a three-layer thermofluid model in fully developed laminar flow problems.* Structural and Multidisciplinary Optimization, 2021. **63**: p. 2789-2809.
32. Yan, et al., *On the non-optimality of tree structures for heat conduction.* INTERNATIONAL JOURNAL OF HEAT AND MASS TRANSFER, 2018.
33. Zhou, J., et al., *Thermal design of microchannel heat sinks using a contour extraction based on topology optimization (CEBTO) method.* International Journal of Heat and Mass Transfer, 2022. **189**: p. 122703.
34. Zeng, S. and P.S. Lee, *Topology optimization of liquid-cooled microchannel heat sinks: An experimental and numerical study.* International Journal of Heat and Mass Transfer, 2019. **142**: p. 118401-.
35. Stolpe, M. and K. Svanberg. *An alternative interpolation scheme for minimum compliance topology optimization*. in *Springer-Verlag*. 2001.
36. Lazarov, B.S. and O. Sigmund, *Filters in topology optimization based on Helmholtz-type differential equations.* International Journal for Numerical Methods in Engineering, 2011. **86**(6): p. 765-781.
37. Xia, Y., et al., *Numerical investigation of microchannel heat sinks with different inlets and outlets based on topology optimization.* Applied Energy, 2023. **330**: p. 120335.

38. Wang, F., B.S. Lazarov, and O. Sigmund, *On projection methods, convergence and robust formulations in topology optimization.* Structural and multidisciplinary optimization, 2011. **43**: p. 767-784.
39. Zuo, K.-T., et al., *Study of key algorithms in topology optimization.* The International Journal of Advanced Manufacturing Technology, 2007. **32**: p. 787-796.
40. Yuan, Y., et al., *Numerical investigation of flow boiling heat transfer in manifold microchannels.* Applied Thermal Engineering, 2022. **217**: p. 119268.
41. Brackbill, J.U., D.B. Kothe, and C. Zemach, *A continuum method for modeling surface tension.* Journal of Computational Physics, 1992. **100**(2): p. 335-354.
42. Jia, Y., et al., *A comparative study of experimental flow boiling heat transfer and pressure drop characteristics in porous-wall microchannel heat sink.* International Journal of Heat and Mass Transfer, 2018. **127**: p. 818-833.
43. Law, M. and P.-S. Lee, *A comparative study of experimental flow boiling heat transfer and pressure characteristics in straight-and oblique-finned microchannels.* International Journal of Heat and Mass Transfer, 2015. **85**: p. 797-810.
44. Zhang, H., et al., *Multiple wall temperature peaks during forced convective heat transfer of supercritical carbon dioxide in tubes.* International Journal of Heat and Mass Transfer, 2021. **172**: p. 121171.
45. Li, Y. and H. Wu, *Experiment investigation on flow boiling heat transfer in a bidirectional counter-flow microchannel heat sink.* International Journal of Heat and Mass Transfer, 2022. **187**: p. 122500.

## S.1 Complete Low-fidelity TO model

### S.1.1 2L model I ($s_1$=1)

For the two-layer TO model, both the fluid channel layer and the solid substrate model are simulated, and they are coupled by an additional source term representing the heat exchange for the two layers in the energy conservation equation. The first two-layer TO model, named 2L model I ($s_1$=1), is proposed by Yan et al. [1], for which the continuity and momentum conservation equations in the fluid channel layer are as follows

$$\nabla \cdot u = 0 \tag{1}$$

$$\frac{6}{7}\rho_{\mathrm{fl}}(u \cdot \nabla)u = -\nabla p + \mu\nabla^2 u + F \tag{2}$$

where $u$, $\rho_{\mathrm{fl}}$, $p$, $\mu$ are the velocity, density, pressure and viscosity of the fluid, respectively. Note that Eq. (2) is different from the original Navier-Stokes equation because the 3D fluid flow in the microchannel is reduced to 2D, and by simplifying the flow in the 3D channel to a parabolic shaped Poiseuille flow, a rescaling constant of 6/7 is added to the momentum equation [1]. $F$ is the Brinkman body force term [2]

$$F = -\alpha(\gamma)u \tag{3}$$

where $\alpha(\gamma)$ is the inverse permeability, which varies in fluid and solid regions due to the continuously changing material distributions during the TO process.

The energy conservation equations in both the fluid channel layer and the substrate layer are based on the later derivation. Originally proposed by [1], the 2L model I takes some assumptions into consideration. A simplification as Poiseuille flow in the channel layer is considered, and the flow direction is set to $x$-direction as shown in Fig. 2. The velocity of the channel layer is

$$\overline{u}_1 = f_1(z)u_1 = \left[1 - \left(\frac{z}{H_{\mathrm{c}}}\right)^2\right]u_1, \quad \overline{u}_2 = 0, \quad \overline{u}_3 = 0 \tag{4}$$

where $\overline{u}_i$ ($i$=1,2,3) is the 3D velocity field, $H_{\mathrm{c}}$ is half of the channel height, and $f_1(z)$ is the non-dimensional velocity distribution function.

Similarly, the non-dimensional temperature $\theta_{\mathrm{c}}$ in the channel layer can be expressed as

$$\theta_{\mathrm{c}} = \frac{T-\overline{T}_{\mathrm{c}}}{T-T_{\mathrm{c}}} = f_2(z) \tag{5}$$

where $T$ is the local temperature, $\overline{T}_{\mathrm{c}}$ is the 3D temperature field in the channel layer, $T_{\mathrm{c}}$ is the bulk mean temperature of the channel layer, $f_2(z)$ is the non-dimensional temperature distribution function of the channel layer. The non-dimensional temperature $\theta_{\mathrm{b}}$ in the substrate layer can be expressed as

$$\theta_{\mathrm{b}} = \frac{T-\overline{T}_{\mathrm{b}}}{T-T_{\mathrm{b}}} = f_3(z) \tag{6}$$

where $T$ is the local temperature, $\overline{T}_{\mathrm{b}}$ is the 3D temperature field, $T_{\mathrm{b}}$ is the bulk mean temperature of the substrate layer, $f_3(z)$ is the non-dimensional temperature distribution function of the substrate layer. The partial derivatives of $\theta_{\mathrm{c}}$ are calculated as

$$\frac{\partial \theta_{\mathrm{c}}}{\partial x} = \frac{\partial}{\partial x}\left(\frac{T-\overline{T}_{\mathrm{c}}}{T-T_{\mathrm{c}}}\right) = 0, \quad \frac{\partial \theta_{\mathrm{c}}}{\partial z} = -\frac{\partial \overline{T}_{\mathrm{c}} / \partial z}{T-T_{\mathrm{c}}} = \frac{\partial f_2(z)}{\partial z} \tag{7}$$

Ignoring the axial heat conduction, the energy equation can be simplified as

$$\rho_{\mathrm{fl}} c_{p\mathrm{fl}} \left( \overline{u}_1 \frac{\partial \overline{T}_c}{\partial x} \right) = \frac{\partial}{\partial z}\left( \lambda_{\mathrm{fl}} \frac{\partial \overline{T}_{\mathrm{c}}}{\partial z} \right) \tag{8}$$

as fluid flow direction is only $x$-direction in Eq. (4). $c_{\mathrm{p,fl}}$ is the fluid heat capacity; $T_{\mathrm{c}}$ is the temperature of the channel layer; $\lambda_{\mathrm{fl}}$ is the thermal conductivity of fluid; At the bottom surface of the channel layer, we have

$$q_{\mathrm{bc}} = h_{\mathrm{c}}\left(T - T_{\mathrm{c}}\right) = -\lambda_{\mathrm{fl}} \left( \frac{\partial \overline{T}_{\mathrm{c}}}{\partial z} \right)_{z=-H_{\mathrm{c}}} \tag{9}$$

where $q_{\mathrm{bc}}$ is the heat flux supplied from the substrate layer to the channel layer, representing the thermal coupling between two layers, $h_{\mathrm{c}}$ is the heat transfer coefficient. Combining Eq. (7) and Eq. (9), we obtain that $h_{\mathrm{c}}/\lambda_{\mathrm{fl}}$ is constant. While the latter is the thermal conductivity of the fluid and therefore constant, $h_{\mathrm{c}}$ is also constant. Then $T$-$T_{\mathrm{c}}$ is constant, $\mathrm{d}T/\mathrm{d}x=\mathrm{d}T_{\mathrm{c}}/\mathrm{d}x$. Combining Eq. (7), we have

$$\frac{\partial \overline{T}_{\mathrm{c}}}{\partial x} = \frac{\mathrm{d}T}{\mathrm{d}x} = \frac{\mathrm{d}T_{\mathrm{c}}}{\mathrm{d}x} \tag{10}$$

Substituting into Eq. (8), we have

$$\frac{\mathrm{d}^2 f_2(z)}{\mathrm{d}z^2} = \frac{\rho_{\mathrm{fl}} c_{p\mathrm{fl}} \mathrm{d}T / \mathrm{d}x}{\lambda_{\mathrm{fl}}(T - T_{\mathrm{c}})} \overline{u}_1 \tag{11}$$

At boundary, we have

$$\overline{T}_{\mathrm{c}} = T \text{ at } z = -H_{\mathrm{c}}, \quad \frac{\partial \overline{T}_{\mathrm{c}}}{\partial z} = 0 \text{ at } z = H_{\mathrm{c}}, \tag{12}$$

Or

$$f_2(z) = 0 \text{ at } z = -H_{\mathrm{c}}, \quad \frac{\mathrm{d}f_2}{\mathrm{d}z} = 0 \text{ at } z = H_{\mathrm{c}}. \tag{13}$$

where $H_{\mathrm{c}}$ is the half height of the fluid channel and solid substrate layer; After integrating Eq. (11) with respect to variable $z$ twice, substituting the velocity distribution in Eq. (4) and boundary condition in Eq. (13), we have

$$f_2(z) = \frac{m}{13}\left[\left(\frac{z}{H_{\mathrm{c}}}\right)^4 - 6\left(\frac{z}{H_{\mathrm{c}}}\right)^2 + 8\left(\frac{z}{H_{\mathrm{c}}}\right) + 13\right] \tag{14}$$

where $m$ is the eigen value introduced by solving the ordinary differential equation Eq. (11).

The bulk mean temperature $T_{\mathrm{c}}$ is defined by

$$T_{\mathrm{c}} = \frac{1}{2H_{\mathrm{c}} u_{\mathrm{m}}} \int_{-H_{\mathrm{c}}}^{H_{\mathrm{c}}} \|\overline{\mathbf{u}}\| \overline{T}_{\mathrm{c}} \, \mathrm{d}z \tag{15}$$

where $u_{\mathrm{m}}$ is the mean velocity of the fluid

$$u_m = \frac{1}{2H_{\mathrm{c}}} \int_{-H_{\mathrm{c}}}^{H_{\mathrm{c}}} \|\overline{\mathbf{u}}\| \mathrm{d}z \tag{16}$$

Replacing 3D temperature $\overline{T}_{\mathrm{c}}$ with Eq. (5)

$$\overline{T}_{\mathrm{c}} = T - f_2(z)(T - T_{\mathrm{c}}) \tag{17}$$

then we have

$$\int_{-H_{\mathrm{c}}}^{H_{\mathrm{c}}} f_1(z) f_2(z) \mathrm{d}z = \frac{4}{3} H_{\mathrm{c}} \tag{18}$$

The non-dimensional temperature distribution function is therefore obtained

$$f_2(z) = \frac{35}{416}\left[\left(\frac{z}{H_{\mathrm{c}}}\right)^4 - 6\left(\frac{z}{H_{\mathrm{c}}}\right)^2 + 8\left(\frac{z}{H_{\mathrm{c}}}\right) + 13\right], \quad z \in [-H_{\mathrm{c}}, H_{\mathrm{c}}] \tag{19}$$

The 3D heat transfer governing equation of the channel layer is

$$\rho_{\mathrm{fl}} c_{p\mathrm{fl}} (\bar{u} \cdot \nabla) \bar{T}_{\mathrm{c}} - \lambda_{\mathrm{fl}} \nabla^2 \bar{T}_{\mathrm{c}} = 0 \tag{20}$$

with boundary conditions

$$\bar{T}_{\mathrm{c}} = T_{\mathrm{in}} \text{ on } \bar{\Gamma}_{\mathrm{in}}, \bar{T}_{\mathrm{c}} = T \text{ on } \bar{\Gamma}_{\mathrm{bc}}, \quad -\lambda_{\mathrm{fl}} \frac{\partial \bar{T}_{\mathrm{c}}}{\partial n} = 0 \text{ on } \bar{\Gamma}_{\mathrm{tc}} \tag{21}$$

where $T_{\mathrm{in}}$ is the inlet temperature at inlet boundary $\bar{\Gamma}_{\mathrm{in}}$, $T$ is the temperature at the interface between the channel layer and the substrate layer $\bar{\Gamma}_{\mathrm{bc}}$, and an adiabatic boundary condition is set to the interface between the channel layer and the top layer $\bar{\Gamma}_{\mathrm{tc}}$.

The weak form of Eq. (20) is:

$$\int_{\Omega_{\mathrm{c}}} \bar{\varphi} \rho_{\mathrm{fl}} c_{p\mathrm{fl}} (\bar{u} \cdot \nabla) \bar{T}_{\mathrm{c}} dV + \int_{\Omega_{\mathrm{c}}} \bar{\varphi} \lambda_{\mathrm{fl}} \nabla^2 \bar{T}_{\mathrm{c}} dV = 0 \tag{22}$$

where $\bar{\varphi} = \bar{\varphi}(x, y, z)$ is the 3D weight function. Considering the temperature field in Eq. (14), we have $\bar{\varphi}(x, y, z) = f_2(z)\varphi(x, y)$, where $\varphi$ is the 2D weight function. Integrating Eq. (22) with some rescaling constants added, we have

$$\int_{\Omega_{\mathrm{c}}} \varphi \left[ \frac{4H_{\mathrm{c}}}{3} \rho_{\mathrm{fl}} c_{p\mathrm{fl}} \left( u \cdot \nabla T_{\mathrm{c}} \right) - \frac{49H_{\mathrm{c}}}{26} \nabla \cdot \left( \lambda_{\mathrm{fl}} \nabla T_{\mathrm{c}} \right) - \frac{35\lambda_{\mathrm{fl}}}{26H_{\mathrm{c}}} \left( T - T_{\mathrm{c}} \right) \right] \mathrm{d}\Omega = 0 \tag{23}$$

The weak form can be reduced as

$$\frac{4H_{\mathrm{c}}}{3} \rho_{\mathrm{fl}} c_{p\mathrm{fl}} \left( u \cdot \nabla T_{\mathrm{c}} \right) - \frac{49H_{\mathrm{c}}}{26} \nabla \cdot \left( \lambda_{\mathrm{fl}} \nabla T_{\mathrm{c}} \right) - \frac{35\lambda_{\mathrm{fl}}}{26H_{\mathrm{c}}} \left( T - T_{\mathrm{c}} \right) = 0 \tag{24}$$

with boundary conditions

$$T = T_{\mathrm{in}} \text{on } \Gamma_{\mathrm{in}}, -\lambda_{\mathrm{fl}} \frac{\partial T}{\partial n} = 0 \text{ on } \Gamma_{\mathrm{t}} \tag{25}$$

Substituting to Eq. (9), we obtain

$$h_{\mathrm{c}} = \frac{35\lambda_{\mathrm{fl}}}{26H_{\mathrm{c}}} \tag{26}$$

The effective heat transfer coefficient $h(\gamma)=26h_{\mathrm{c}}/35$ is introduced and Eq. (24) is therefore reformulated as

$$\frac{2}{3} u \cdot \nabla \left( \rho c_{p\mathrm{fl}} T_{\mathrm{c}} \right) = \frac{49}{52} \nabla \cdot \left( \lambda(\gamma) \nabla T_{\mathrm{c}} \right) + \frac{h(\gamma)\left( T_{\mathrm{b}} - T_{\mathrm{c}} \right)}{2H_{\mathrm{c}}} \tag{27}$$

The linear temperature distribution in the substrate layer with respect to $z$-direction is considered and can be written as

$$\overline{T}_b = T + \frac{q_{\mathrm{bc}} H_b}{k_b}\left(1 - \frac{z}{H_b}\right) \tag{28}$$

where $H_{\mathrm{b}}$ is the half height of the solid substrate layer; Using the same weak-form-elimination process, the obtained governing equation is

$$\nabla \cdot \left(\frac{\lambda_{\mathrm{s}}}{2} \nabla T_{\mathrm{b}}\right) + \frac{q_{\mathrm{in}}}{2H_{\mathrm{b}}} - \frac{h(\gamma)(T_{\mathrm{b}} - T_{\mathrm{c}})}{2H_{\mathrm{b}}} = 0 \tag{29}$$

Where $\lambda_{\mathrm{fl}}$ is the thermal conductivity of solid. Therefore, the reduced heat transfer equations Eq. (27) and Eq. (29) are derived for 2L model I. $q_{\mathrm{in}}$ is the heat flux applied at the bottom of the solid substrate layer. Considering the concept of thermal circuits, the thermal resistance of the solid substrate and channel layers in the y-direction is connected in series, thus the heat transfer coefficient between the two layers $h(\gamma)$ is calculated by

$$h(\gamma) = \frac{h_{\mathrm{b}} h_{\mathrm{c}}}{(h_{\mathrm{b}} + h_{\mathrm{c}})} \tag{30}$$

where $h_{\mathrm{b}}$ is the reciprocal of the conductive thermal resistance $R_{\mathrm{b}}$ in the solid substrate layer. Here, $R_{\mathrm{b}}$ from the middle plane of the solid layer to the interface between the solid layer and the fluid layer can be calculated as:

$$R_{\mathrm{b}} = \frac{1}{h_{\mathrm{b}}} = \frac{H_{\mathrm{b}}}{\lambda_{\mathrm{s}}} \tag{31}$$

Therefore,

$$h_{\mathrm{b}} = \frac{\lambda_{\mathrm{s}}}{H_{\mathrm{b}}} \tag{32}$$

$h_{\mathrm{c}}$ is the reciprocal of the convective thermal resistance in the fluid channel layer [1]. Similarly, the convective thermal resistance $R_{\mathrm{c}}$ can be expressed as

$$R_{\mathrm{c}} = \frac{1}{h_{\mathrm{c}}} = \frac{H_{\mathrm{c}}}{\lambda(\gamma)} \tag{33}$$

where $\lambda(\gamma)$ is the effective thermal conductivity

$$\lambda(\gamma) = \begin{cases} \lambda_{\mathrm{fl}} & \gamma \in \Omega_{\mathrm{fl}} \\ \lambda_{\mathrm{s}} & \gamma \in \Omega_{\mathrm{s}} \end{cases} \tag{34}$$

Thus, $h_{\mathrm{c}}$ is different in $\Omega_{\mathrm{fl}}$ and $\Omega_{\mathrm{s}}$

$$h_c = \begin{cases} h_{fl} = \dfrac{35\lambda_{fl}}{26H_c} & \gamma \in \Omega_{fl} \\ h_s = \dfrac{35\lambda_s}{26H_c} & \gamma \in \Omega_s \end{cases} \tag{35}$$

In the present study, the inverse permeability $\alpha(\gamma)$ in Eq. (3) is interpolated by the Darcy interpolation function [3]

$$\alpha(\gamma) = \alpha_{fl} + (\alpha_s - \alpha_{fl})\frac{1-\gamma}{1+q_f\gamma} \tag{36}$$

where $\alpha_{fl}$ is the inverse permeability in the fluid domain

$$\alpha_{fl} = \frac{5\mu}{2H_c^2} \tag{37}$$

$\alpha_s$ is the inverse permeability in the solid domain [3]

$$\alpha_s = \frac{5\mu}{2\left(\dfrac{H_c}{100}\right)^2} \tag{38}$$

$q_f$ is the penalization parameter that is adopted to regulate the convexity of the interpolation function. In the present study, $q_f$ is set as 1000, 500, 100 and 20 in different TO steps. In each step, the TO solver runs 50 times or meets the tolerance of 0.01.

**S.1.2 2L model II ($s_1$=2)**

The continuity and the momentum conservation equations of the second kind of two-layer model, named 2L model II ($s_1$=2), are similar to Eqs. (3) and (4) [4]. The difference between the two 2L models is that the rescaling constant is not included in the governing conservation equations of 2L model II, which are shown as follows

$$\nabla \cdot u = 0 \tag{39}$$

$$\rho_{fl}(u \cdot \nabla)u = -\nabla p + \mu\nabla^2 u + F \tag{40}$$

$$u \cdot \nabla\left(\rho_{fl} c_{p,fl} T_c\right) = \nabla \cdot \left(\lambda(\gamma)\nabla T_c\right) + \frac{h(\gamma)(T_b - T_c)}{2H_c} \tag{41}$$

$$\nabla \cdot (\lambda_s \nabla T_s) + \frac{q_{in}}{2H_b} - \frac{h(\gamma)(T_b - T_c)}{2H_b} = 0 \tag{42}$$

In the 2L model II, the energy equation of fluid in the fluid channel layer is written

as

$$u\cdot\nabla\left(\rho_{\mathrm{fl}}c_{p,\mathrm{fl}}T_{\mathrm{c}}\right)=\nabla\cdot\left(\lambda_{\mathrm{fl}}\nabla T_{\mathrm{c}}\right)+\frac{h_{\mathrm{fl}}\left(T_{\mathrm{b}}-T_{\mathrm{c}}\right)}{2H_{\mathrm{c}}} \tag{43}$$

And the energy equation of the solid in the fluid channel is written as

$$\nabla\cdot\left(\lambda_{\mathrm{s}}\nabla T_{\mathrm{fl}}\right)+\frac{h_{\mathrm{s}}\left(T_{\mathrm{b}}-T_{\mathrm{c}}\right)}{2H_{\mathrm{c}}}=0 \tag{44}$$

In the solid substrate channel layer, heat is either conducted into fins or convectively transferred to the liquid. In domains beneath fluid domains of the fluid channel layer:

$$\nabla\cdot\left(\lambda_{\mathrm{s}}\nabla T_{\mathrm{s}}\right)+\frac{q_{\mathrm{in}}}{2H_{\mathrm{b}}}-\frac{h_{\mathrm{fl}}\left(T_{\mathrm{b}}-T_{\mathrm{c}}\right)}{2H_{\mathrm{b}}}=0 \tag{45}$$

In domains beneath solid domains of the fluid channel layer:

$$\nabla\cdot\left(\lambda_{\mathrm{s}}\nabla T_{\mathrm{s}}\right)+\frac{q_{\mathrm{in}}}{2H_{\mathrm{b}}}-\frac{h_{\mathrm{s}}\left(T_{\mathrm{b}}-T_{\mathrm{c}}\right)}{2H_{\mathrm{b}}}=0 \tag{46}$$

where $h_{\mathrm{fl}}$ represents the convection heat transfer resistance between base surface without fin and fluid, which is calculated as

$$h_{\mathrm{fl}}=\frac{q_{\mathrm{unfin}}}{A_{\mathrm{unfin}}(T_{\mathrm{unfin}}-T_{fl})} \tag{47}$$

where $q_{\mathrm{unfin}}$ is the heat transfer rate from area of solid substrate without fin to fluid, $A_{\mathrm{unfin}}$ is the unfinned area of solid substrate, $T_{\mathrm{unfin}}$ is the average wall temperature of the unfinned area of solid substrate, and $T_{\mathrm{fl}}$ is the volume weighted average temperature of fluid in the channel domain. $h_{\mathrm{s}}$ reflects the conduction heat transfer resistance between the finned base solid substrate surface and the fins, so it is called nominal convection heat transfer coefficient, which can be calculated as:

$$h_{\mathrm{s}}=\frac{\lambda_{\mathrm{s}}}{H_{\mathrm{c}}} \tag{48}$$

Therefore, the energy conservation equations of 2L model II are shown as follows

$$u\cdot\nabla\left(\rho_{\mathrm{fl}}c_{p,\mathrm{fl}}T_{\mathrm{c}}\right)=\nabla\cdot\left(\lambda(\gamma)\nabla T_{\mathrm{c}}\right)+\frac{h(\gamma)\left(T_{\mathrm{b}}-T_{\mathrm{c}}\right)}{2H_{\mathrm{c}}} \tag{49}$$

$$\nabla\cdot\left(\lambda_{s}\nabla T_{s}\right)+\frac{Q_{in}}{2H_{b}}-\frac{h(\gamma)\left(T_{b}-T_{c}\right)}{2H_{b}}=0 \tag{50}$$

The thermal conductivity and nominal heat transfer coefficient should satisfy that

$$\lambda(\gamma)=\begin{cases}\lambda_{fl} & \gamma\in\Omega_{fl}\\ \lambda_{s} & \gamma\in\Omega_{s}\end{cases}\qquad h(\gamma)=\begin{cases}h_{fl} & \gamma\in\Omega_{fl}\\ h_{s} & \gamma\in\Omega_{s}\end{cases} \tag{51}$$

### S.1.3 1L model ($s_1$=3)

The continuity and momentum conservation equations of the incompressible steady fluid flow for the 1L model are given as:

$$\nabla\cdot u=0 \tag{52}$$

$$\rho_{fl}\left(u\cdot\nabla\right)u=-\nabla p+\mu\nabla^{2}u+F \tag{53}$$

where $F$ is the Brinkman body force term and the inverse permeability is $\alpha_{max}q(1-\gamma)/(1+\gamma)$. There are different governing equations for the thermal field in the fluid and solid domains. When the heat source is only placed in the solid domain, the energy equation can be expressed as:

$$\rho_{fl}c_{pfl}\left(u\cdot\nabla\right)T_{c}=\nabla\cdot\left(\lambda_{fl}\nabla T_{c}\right),\quad \text{(in fluid domains)} \tag{54}$$

$$0=\nabla\cdot\left(\lambda_{s}\nabla T_{c}\right)+\frac{q_{in}}{2H_{c}},\quad \text{(in solid domains)} \tag{55}$$

The two equations, Eq. (54) and Eq. (55), are unified into the following single equation which is applicable to both the fluid and solid domains:

$$\gamma\rho_{fl}c_{pfl}\left(u\cdot\nabla\right)T_{c}=[(1-\gamma)\frac{\lambda_{s}}{\lambda_{f}}+\gamma]\nabla^{2}T_{c}+(1-\gamma)\frac{q_{in}}{2H_{c}} \tag{56}$$

### S.1.4 Model implementation

Based on the governing equations with different $s_1$, the low-fidelity optimization problem for the microchannel heat sink is defined as

$$\begin{aligned}&\min:\ T_{max}\\ &\text{s.t.}:\ \begin{cases}\text{governing equations}\\ \dfrac{\int_{\Omega_{d}}\gamma dA}{A_{\Omega_{d}}}=f_{c}\\ 0\leqslant\gamma\leqslant 1\end{cases}\end{aligned} \tag{57}$$

where $T_{max}$ is the maximum temperature of the heat sink, and the minimum $T_{max}$ corresponds to the highest heat transfer performance. $A_{\Omega d}$ is the area of the design domain $\Omega_d$. $f_c$ is the fluid volume fraction constraint.

To obtain a differentiable maximum operator, the maximum temperature in the computational domain is calculated by a generalized $p$-norm

$$T_{\max} = \left( \frac{1}{N} \sum_N T_i^p \right)^{\frac{1}{p}} \tag{58}$$

where $N$ is the total number of nodes in the computational domain and $p$ is the temperature form parameter set as 10 [5, 6].

The effective thermal conductivity $\lambda(\gamma)$ is interpolated by the rational approximation of material properties (RAMP) function [7]

$$\lambda(\gamma) = \frac{\gamma\left[\lambda_{\mathrm{fl}}\left(1+q_{\mathrm{k}}\right)-\lambda_{\mathrm{s}}\right]+\lambda_{\mathrm{s}}}{1+q_{\mathrm{k}}\gamma} \tag{59}$$

where $q_k$ is the penalization parameter set as 10, 50 250 and 1000 in different TO steps.

To avoid the checkerboard problems, the Helmholtz based on partial differential equation (PDE) density filter is adopted [8]

$$-r^2\nabla^2\gamma_{\mathrm{f}} + \gamma_{\mathrm{f}} = \gamma \tag{60}$$

where $\gamma_f$ is the filtered design variable, and $r$ is the filter radius. In the present study, $r$ is set to twice the cell size [9].

The density filter may lead to the intermediate density area at the fluid-solid interface. To obtain clearer fluid-solid interface, the hyperbolic tangent projection is employed [10]

$$\gamma_{\mathrm{p}} = \frac{\tanh\left(\beta\left(\gamma_{\mathrm{f}}-\gamma_{\mathrm{c}}\right)\right)+\tanh\left(\beta\gamma_{\mathrm{c}}\right)}{\tanh\left(\beta\left(1-\gamma_{\mathrm{c}}\right)\right)+\tanh\left(\beta\gamma_{\mathrm{c}}\right)} \tag{61}$$

where $\gamma_p$ is the projected design variable. $\gamma_p$ is the projection point which is set to 0.5. $\beta$ is the projection slope which is set to 8.

## S.2 Complete high-fidelity phase-change model

### S.2.1 Governing equations

For the flow boiling process, conservation equations of mass, momentum, energy, and volume fraction of vapor are solved

$$\frac{\partial(\rho)}{\partial t} + \nabla\cdot(\rho\mathbf{u}) = 0 \tag{62}$$

$$\frac{\partial(\rho\mathbf{u})}{\partial t}+\nabla\cdot(\rho\mathbf{u}\mathbf{u})=-\nabla p+\nabla\cdot[\mu(\nabla\mathbf{u}+\nabla\mathbf{u}^{\mathrm{T}})]+\rho\mathbf{g}+\mathbf{F} \tag{63}$$

$$\frac{\partial(\rho c_{\mathrm{p}}T)}{\partial t}+\nabla\cdot(\rho c_{\mathrm{p}}T\mathbf{u})=\nabla\cdot(\lambda\nabla T)+S_{\mathrm{E}} \tag{64}$$

$$\frac{\partial(C_k\rho_k)}{\partial t}+\nabla\cdot(C_k\rho_k\mathbf{u})=\dot{m}_k \tag{65}$$

where $t$ is the time, $p$ is the pressure, and $\mathbf{u}$ is the velocity. $\rho$ and $\mu$ are volume averaged density and dynamic viscosity of the two phases. $\mathbf{g}$ is the gravitational acceleration. $T$ is the temperature, $c_{\mathrm{p}}$ is the specific heat capacity, and $\lambda$ is the thermal conductivity. The body force term $\mathbf{F}$ is used to characterize the effect of surface tension force which is converted from surface force to body force by the continuum surface force (CSF) model [11]

$$\mathbf{F}=2\sigma\rho\frac{\nabla\cdot(\frac{\mathbf{n}}{|\mathbf{n}|})\nabla(C_k)}{\rho_1+\rho_2} \tag{66}$$

where $\sigma$ is the surface tension coefficient. $\rho_1$ and $\rho_2$ are the density of the primary phase and secondary phase, respectively. $C_k$ is the volume fraction of $k$th phase. In the present study, $C_1$ and $C_2$ are for liquid as the primary phase and vapor as the secondary phase, respectively. The interface normal direction $\mathbf{n}$ is defined as the gradient of $C_k$

$$\mathbf{n}=\nabla C_k \tag{67}$$

Note that $\mu$, $\rho$ and $\lambda$ in conservation equations Eqs. (61), (62) and (63) are volume averaged value calculated by linear interpolation, and the $c_{\mathrm{p}}$ is mass averaged

$$\rho=C_1\rho_1+C_2\rho_2 \quad \mu=C_1\mu_1+C_2\mu_2 \quad \lambda=C_1\lambda_1+C_2\lambda_2 \quad c_{\mathrm{p}}=\frac{C_1\rho_1c_{\mathrm{p},1}+C_2\rho_2c_{\mathrm{p},2}}{C_1\rho_1+C_2\rho_2} \tag{68}$$

$S_{\mathrm{E}}$ in the energy conservation equation is the energy source term caused by the phase change, which can be calculated based on the boiling rate $\dot{m}$

$$S_{\mathrm{E}}=h_{\mathrm{fg}}\dot{m} \tag{69}$$

where $h_{\mathrm{fg}}$ is the latent heat. For the boiling process, the relationship among liquid reduction rate $\dot{m}_2$, vapor generation rate $\dot{m}_1$, and the boiling rate in Eq. (69)

$$\dot{m} = -\dot{m}_1 = \dot{m}_2 \tag{70}$$

### S.2.2 Phase interface recognition model

Fig. A1 schematically shows a local liquid-vapor distribution and the values of volume fraction of vapor in each computational cell. Here, the scheme proposed by Rattner and Garimella [12] is adopted to identify the phase interface. For every two neighbor cells, called a cell-pair, if $C_l$ of one cell is larger than a prescribed value $\tilde{C}_1$ and that of the other cell is lower than $\tilde{C}_1$, then the phase interface is considered to be located in this cell-pair. For the boiling processes considered, in the cell with higher $C_l$ of a phase interface cell-pair, the source terms in the energy conservation equations of and volume fraction will be calculated. All the cell-pairs of phase interface in the computational domain can be found out based on the above standard, as shown in Fig. A1. It is worth mentioning that for a cell directly adjacent to the solid wall, if its $C_l$ is higher than the prescribed value $\tilde{C}_1$, it is directly treated as an interfacial cell which allows the source terms to be calculated. This is because by such scheme the nucleation at the solid wall can be implemented. In the present study, the value of $\tilde{C}_1$ is set as 0.8, using which it has been demonstrated that the phase interface can be well described.

### S.2.3 The phase change model

In the present study, a heat-conduction-based method is employed. The basic idea is that at the phase interface where phase change possibly occurs, the local temperature should converge to the local saturation temperature $T_{sat}$. The energy source term $S_E$ is thus determined to satisfy the above requirement, and the corresponding scheme is schematically explained by one-dimensional situation as shown in Fig. A2, which can be easily extended to 2D and 3D processes. Fig. A2 shows the nodes (W, P, E, with distance between them as $\delta x$), the control volume of P (the gray region with width of $\Delta x$) and the interfaces of the control volume (the dashed lines $w$ and $e$) based on the finite volume method. The discretized form of the energy conservation equation for node P is as follow

$$a_P T_P = a_E T_E + a_W T_W + b \tag{71}$$

For second order upwind difference scheme adopted in this model, the coefficient $a$ and the source term $b$ are calculated as follows

$$a_{\mathrm{E}} = D_e + [|-F_e, 0|],\quad a_{\mathrm{W}} = D_w + [|-F_w, 0|],\quad a_{\mathrm{P}} = a_{\mathrm{E}} + a_{\mathrm{W}} + a_{\mathrm{P}}^0 \tag{72}$$

$$b = S_{\mathrm{E}}^n + a_{\mathrm{P}}^0 T_{\mathrm{P}}^0 \tag{73}$$

$$D_e = \frac{\lambda\left|\overrightarrow{A_e}\right|}{\delta x_e},\quad F_e = \rho c_{\mathrm{p}} \overrightarrow{U_e} \cdot \overrightarrow{A_e},\quad a_{\mathrm{P}}^0 = \frac{\rho c_{\mathrm{p}} V_{\mathrm{cell}}}{\Delta t} \tag{74}$$

where $D$ is the reciprocal of diffusion resistance, and $F$ is the flow rate on the interface of the control volume. $\vec{A}$ is the area of the interface, $\vec{U}$ is the velocity on the interface of the control volume and $V_{\mathrm{cell}}$ is the volume of the control volume. The superscript 0 and $n$ mean the last time and current step, respectively. The operator "[| |]" is equal to the maximum value of all the parameters in the operator.

According to above definition, Eq. (71) can be deformed into following equation

$$(a_{\mathrm{E}} + a_{\mathrm{W}} + a_{\mathrm{P}}^0) T_{\mathrm{P}} = a_{\mathrm{E}} T_{\mathrm{E}} + a_{\mathrm{W}} T_{\mathrm{W}} + a_{\mathrm{P}}^0 T_{\mathrm{P}}^0 + S_{\mathrm{E}}^n \tag{75}$$

As mentioned, the basic idea of our phase change model is to converge the local temperature to the saturated value, and thus the discretization of energy equation in the next time step can be written as follow by assuming that the saturation temperature is obtained

$$(a_{\mathrm{E}} + a_{\mathrm{W}} + a_{\mathrm{P}}^0) T_{\mathrm{sat}} = a_{\mathrm{E}} T_{\mathrm{E}} + a_{\mathrm{W}} T_{\mathrm{W}} + a_{\mathrm{P}}^0 T_{\mathrm{P}} + S_{\mathrm{E}}^{n+1} \tag{76}$$

where the superscript $n$+1 means the next time step. With explicit time scheme adopted, the difference of $S_{\mathrm{E}}$ can be obtained by subtracting the above two equations

$$S_{\mathrm{E}}^{n+1} - S_{\mathrm{E}}^n = (a_{\mathrm{E}} + a_{\mathrm{W}})(T_{\mathrm{sat}} - T_{\mathrm{P}}) + a_{\mathrm{P}}^0 (T_{\mathrm{sat}} - 2T_{\mathrm{P}} + T_{\mathrm{P}}^0) \tag{77}$$

Eq. (77) provides the iteration formula for the energy source term $S_{\mathrm{E}}$. According to this formula, $S_{\mathrm{E}}$ at the next time step can be calculated by $S_{\mathrm{E}}$, $T_{\mathrm{P}}$ and other parameters of the current time step. It should be noted that Eq. (77) includes the $T_{\mathrm{p}}$ at next time step as a reclusive formula, thus $S_{\mathrm{E}}^{n+1}$ is determined with several iterations, and the value of $T_{\mathrm{P}}$ and $S_{\mathrm{E}}^{n+1}$ are updated in every iteration. To prevent extremely large $\dot{m}$, the Hertz-Knudsen equation is employed [13].

$$\frac{S_{\mathrm{E}}^{n+1}}{h_{\mathrm{fg}}} \le \dot{m}_{n+1,\lim} = \sigma_{\mathrm{c}} \sqrt{\frac{M}{2\pi R}} \left(\frac{p}{\sqrt{T}} - \frac{p_{\mathrm{sat}}(T)}{\sqrt{T}}\right) A_{\mathrm{if}} \tag{78}$$

where $A_{\mathrm{if}}$ is the area of the phase interface in current cell. $p_{\mathrm{sat}}$ is the saturation pressure. $\sigma_{\mathrm{c}}$ is the relaxation parameter introduced by molecular collisions and other factors. $M$ is the molecular mass and $R$ is the gas constant. After $S_{\mathrm{E}}^{n+1}$ is obtained, $\dot{m}$ can be determined based on Eq. (70).

For 3D model, the above 1D reclusive formula is extended

$$S_{\mathrm{E}}^{n+1} - S_{\mathrm{E}}^{n} = (a_{\mathrm{E}} + a_{\mathrm{W}} + a_{\mathrm{N}} + a_{\mathrm{S}} + a_{\mathrm{T}} + a_{\mathrm{B}})(T_{\mathrm{sat}} - T_{\mathrm{P}}) + a_{\mathrm{P}}^{0}(T_{\mathrm{sat}} - 2T_{\mathrm{P}} + T_{\mathrm{P}}^{0}) \tag{79}$$

where the subscripts N,S,T and B represent the south, north, top and bottom nodes, respectively.

For water as coolant in the present study, the Antoine equation is adopted to calculate the saturation temperature

$$T_{\mathrm{sat}} = \frac{3826.36}{9.3876 - \ln p_{\mathrm{sat}}} - 227.68 \tag{80}$$

The water latent heat is as follow

$$h_{\mathrm{fg}} = -0.002719 \times T^{2} + 2.036 \times T + 2499 - 4.18 \times T \tag{81}$$

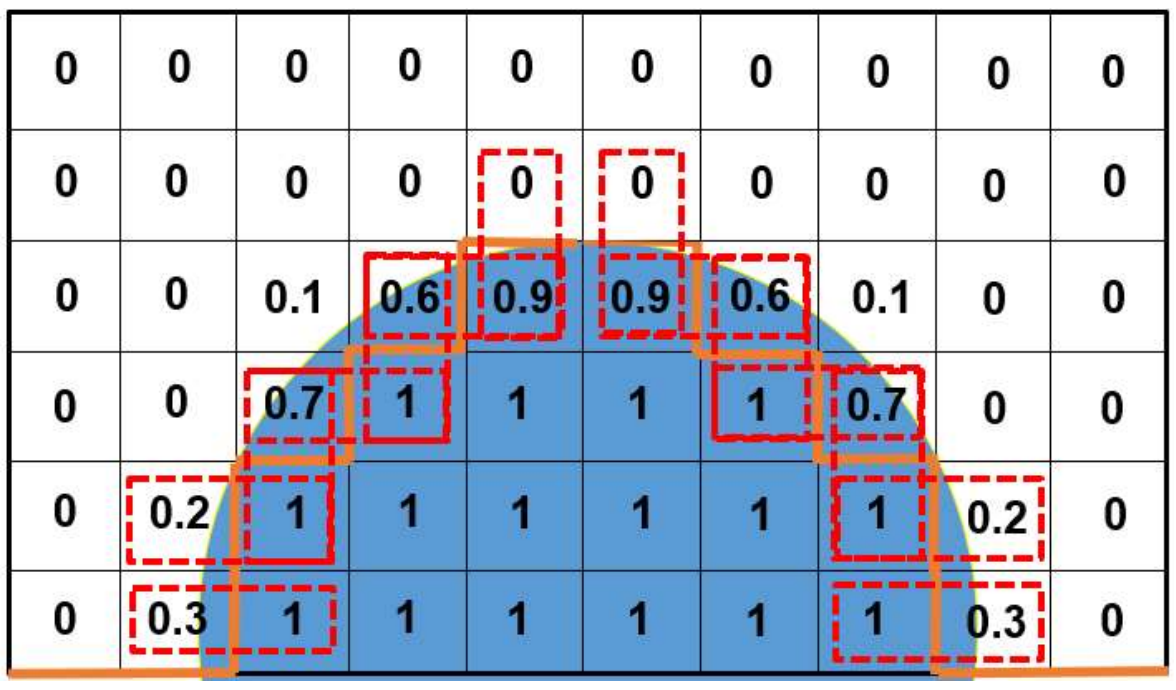

Fig. A1 Schematic of the liquid-water vapor distribution and values of volume of fraction of the liquid phase. The read box stands for the cell-pairs in which the phase interface is located. The orange line is the phase interface identified by the model.

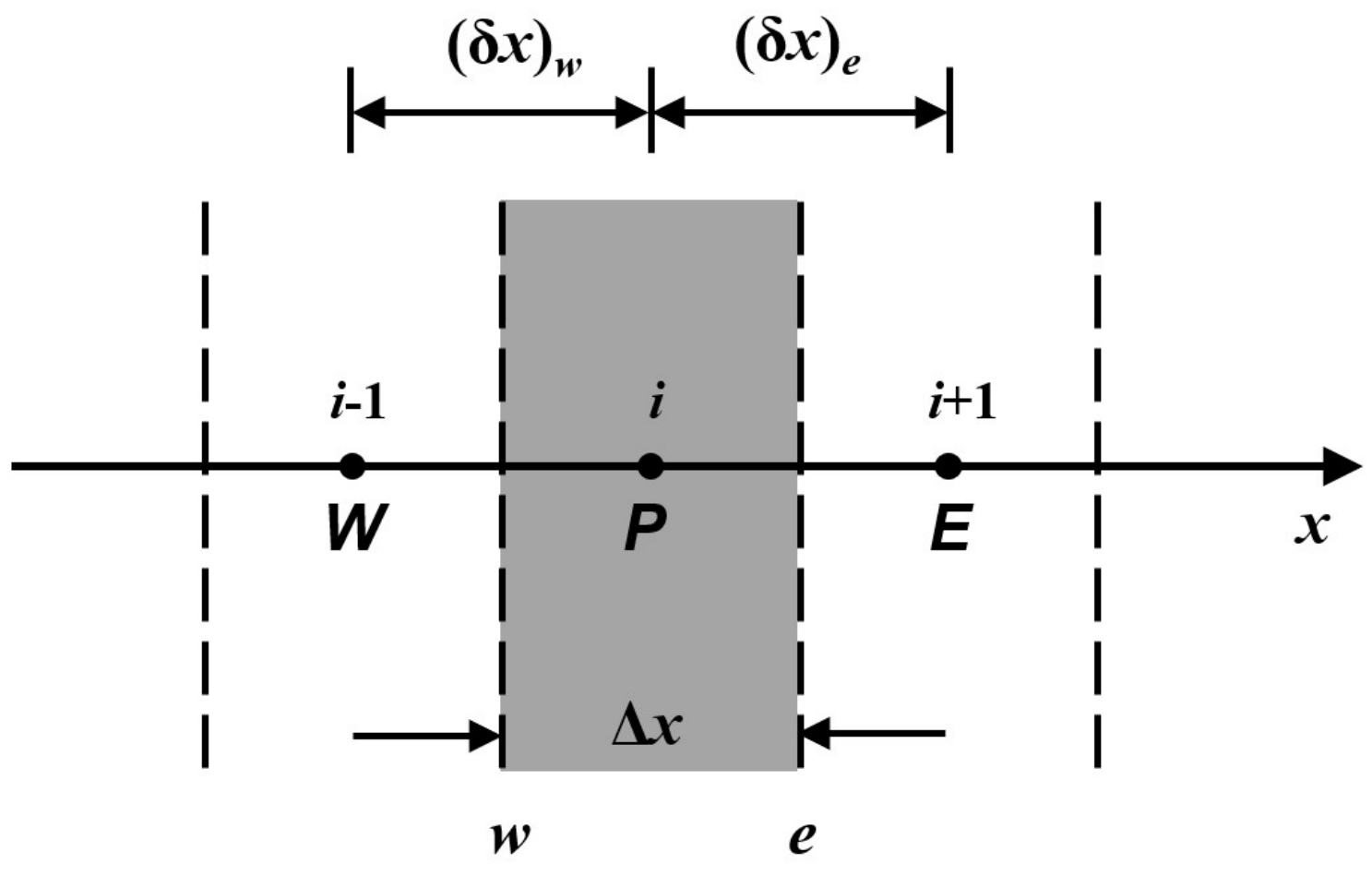

Fig. A2 Schematic of node, interface and control volume based on the finite volume method (one-dimension).

**S.3 Validation of the phase-change model**

A classic film boiling process, as shown in Fig. A3, is adopted to verify the present phase change heat transfer model. The computational domain is a rectangular with size of 15× 15× 45 mm. A high temperature of 378.15 K (greater than the saturation temperature $T_{sat}$ 373.15 K) is set on the square bottom wall, and the outflow boundary condition is employed at the top surface. The symmetric boundary condition is used for the rest of the four boundaries. The absolute pressure, temperature and the velocity of the whole computational domain are initially set as 101325 Pa, 373.15 K and 0 m $s^{-1}$, respectively. According to experimental and theoretical studies in the literature, the bubbles generated during the film boiling process are distributed uniformly on the heated bottom wall, and the critical wavelength (the bubbles spacing during film boiling) is $l_c=2\pi(3\sigma/(\rho_l-\rho_g)g)^{1/2}$. The bubbles will be generated periodically on the substrate with constant intervals as $l_c$. In order to reproduce this regular phenomenon, in our simulation the side length of the heated bottom wall is set as $l_c$ /2.

With the above settings, vapor will be generated from the bottom surface and rises up under the effects of buoyancy, as shown by Fig. A3(a) in which the time evolution of the vapor interface in the domain is displayed. It can be found that the bubbles are generated and detached on each corner in turn. The averaged *Nu* of the heating surface is further calculated based on the temperature field obtained from the simulation

$$Nu = \frac{Ql_c}{2A_{bot}\lambda(T_{ave} - T_{sat})} \tag{82}$$

where $Q$ is the total heat flux on the bottom surface, and $A_{bot}$ is the area of bottom surface. As shown in Fig. A3(b), $Nu$ increases when the bubble starts to generate on the heating surface, and decreases when the bubble starts to detach from the heating surface. Berenson [14] proposed the following correlation for $Nu$

$$Nu = 0.425(\frac{\rho_g(\rho_l - \rho_g)gh_{fg}}{\lambda_g \mu_g (T_{wall} - T_{sat})})^{1/4} l_c^{3/4} \tag{83}$$

The value of $Nu$ calculated is also plotted in Fig. A3(b) as the black line, which is a constant as 280. The averaged value of $Nu$ based on our simulation is 264, in good agreement with the value predicted by E.A2, with relative error about 5.7%.

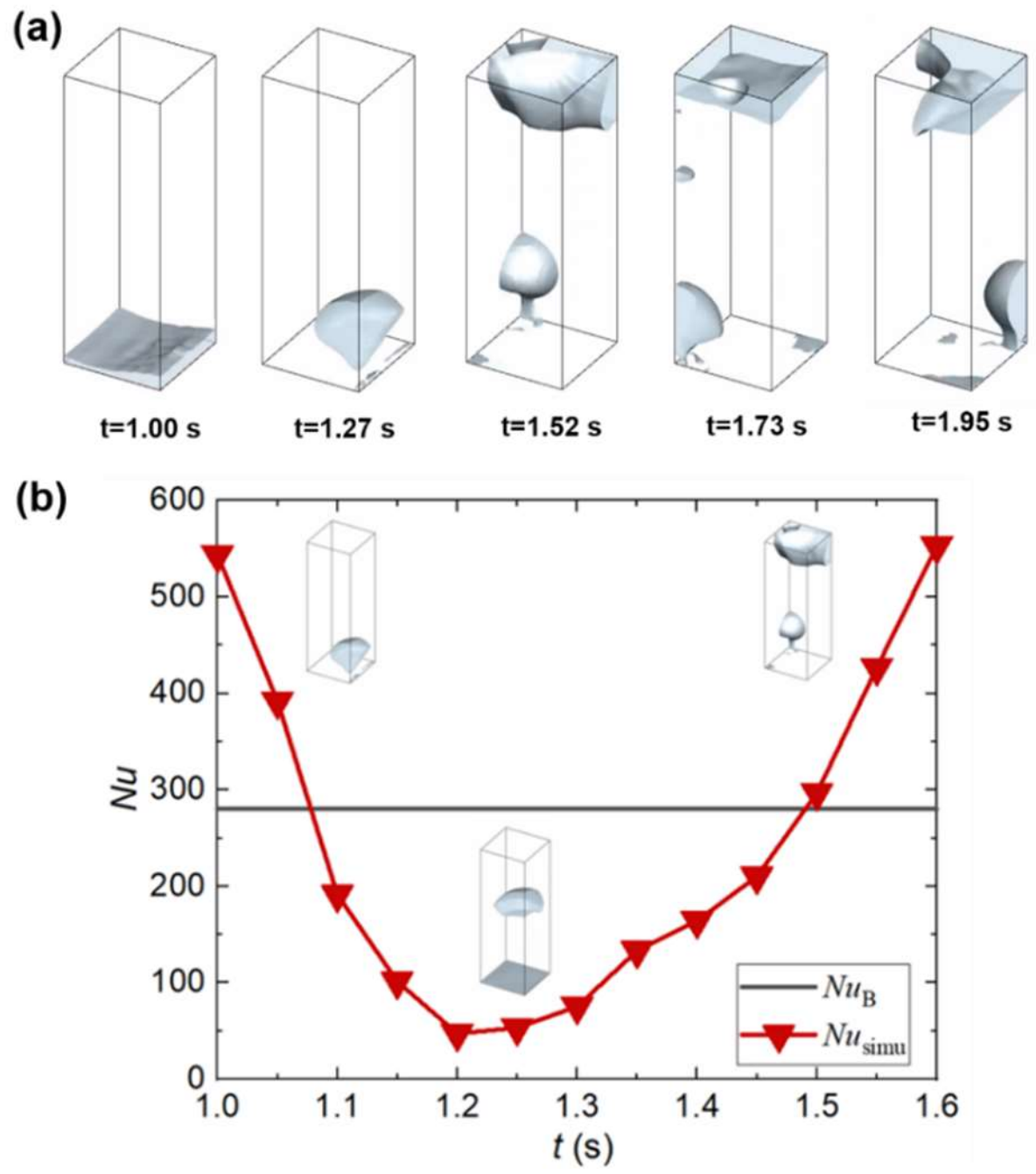

Fig. A3 Simulation results for the validation of the phase change model. (a) Growth and detachment process of bubbles in film boiling. (b) The average value of $Nu$ of the heating surface.

## S.4 Supplementary material of results and discussion

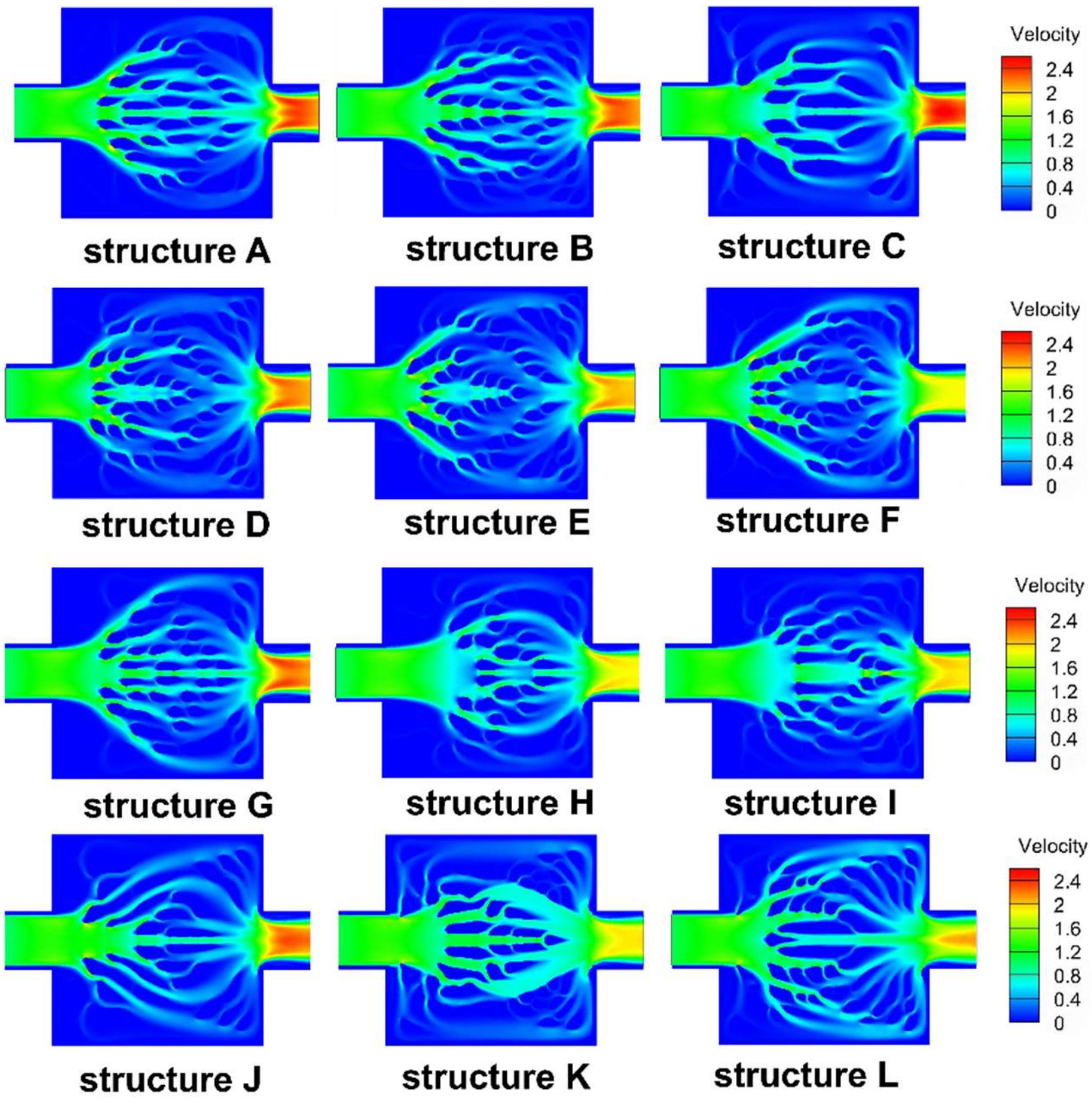

Fig. A4 Velocity fields of different structures.

## References

1. Yan, S., et al., *Topology optimization of microchannel heat sinks using a two-layer model.* International Journal of Heat and Mass Transfer, 2019. **143**(NOV.): p. 118462.1-118462.16.
2. Borrvall, T. and J. Petersson, *Topology optimization of fluids in Stokes flow.* International journal for numerical methods in fluids, 2003. **41**(1): p. 77-107.
3. Zeng, S. and P.S. Lee, *Topology optimization of liquid-cooled microchannel heat sinks: An experimental and numerical study.* International Journal of Heat and Mass Transfer, 2019. **142**: p. 118401-.
4. Zeng, S. and P.S. Lee, *Topology optimization of liquid-cooled microchannel heat sinks: An experimental and numerical study.* International Journal of Heat and Mass Transfer, 2019. **142**: p. 118401.
5. Yan, et al., *On the non-optimality of tree structures for heat conduction.* INTERNATIONAL JOURNAL OF HEAT AND MASS TRANSFER, 2018.

6. Zhou, J., et al., *Thermal design of microchannel heat sinks using a contour extraction based on topology optimization (CEBTO) method.* International Journal of Heat and Mass Transfer, 2022. **189**: p. 122703.
7. Stolpe, M. and K. Svanberg. *An alternative interpolation scheme for minimum compliance topology optimization*. in *Springer-Verlag*. 2001.
8. Lazarov, B.S. and O. Sigmund, *Filters in topology optimization based on Helmholtz-type differential equations.* International Journal for Numerical Methods in Engineering, 2011. **86**(6): p. 765-781.
9. Xia, Y., et al., *Numerical investigation of microchannel heat sinks with different inlets and outlets based on topology optimization.* Applied Energy, 2023. **330**: p. 120335.
10. Wang, F., B.S. Lazarov, and O. Sigmund, *On projection methods, convergence and robust formulations in topology optimization.* Structural and multidisciplinary optimization, 2011. **43**: p. 767-784.
11. Brackbill, J.U., D.B. Kothe, and C. Zemach, *A continuum method for modeling surface tension.* Journal of computational physics, 1992. **100**(2): p. 335-354.
12. Rattner, A.S. and S. Garimella, *Simple mechanistically consistent formulation for volume-of-fluid based computations of condensing flows.* Journal of heat transfer, 2014. **136**(7): p. 071501.
13. Schrage, R.W., *A theoretical study of interphase mass transfer*. 1953: Columbia University Press.
14. Berenson, P., *Experiments on pool-boiling heat transfer.* International Journal of Heat and Mass Transfer, 1962. **5**(10): p. 985-999.